%% file: main_text_converted.tex
\documentclass[reprint,amsmath,amssymb,aps]{revtex4-2}

\usepackage{graphicx}

\usepackage{siunitx}
\usepackage{dcolumn}
\usepackage{bm}
\usepackage{color}
\usepackage{xcolor}
\usepackage{needspace}
\DeclareUnicodeCharacter{2212}{-}
\usepackage[colorlinks,bookmarks=true,citecolor=blue,linkcolor=red,urlcolor=blue]{hyperref}

\usepackage{subfigure}

\begin{document}

\title{Magnons reveal topology and dynamics of a skyrmion crystal}

\author{R. Ayache$^{1,\dagger}$}
\author{N. Chakraborty$^{2,\dagger}$}
\author{M. Kuiri$^{1,3,\dagger}$}
\author{Q. Benichou$^{1}$}
\author{A. Lacerda-Santos$^{1}$}
\author{L. Seyve$^{1}$}
\author{H. Chakraborti$^{1,4}$}
\author{L. Pugliese$^{1}$}
\author{K. Watanabe$^{5}$}
\author{T. Taniguchi$^{5}$}
\author{C. Gorini$^{1}$}
\author{R. Moessner$^{6}$}
\author{B. Doucot$^{7}$}
\author{P. Roulleau$^{1,*}$}

\affiliation{$^{1}$SPEC, CEA, CNRS, Université Paris-Saclay, CEA Saclay, 91191 Gif-sur-Yvette, France}

\affiliation{$^{2}$TCM Group, Cavendish Laboratory, University of Cambridge, Cambridge CB3 0HE, United Kingdom}

\affiliation{$^{3}$Department of Physics, Birla Institute of Technology and Science, Pilani Hyderabad Campus, Telangana 500078, India}

\affiliation{$^{4}$Laboratory of Atomic and Solid State Physics, Cornell University, Ithaca, NY 14850, USA}

\affiliation{$^{5}$National Institute for Materials Science, Tsukuba 305-0044, Japan}

\affiliation{$^{6}$Max Planck Institute for the Physics of Complex Systems, Dresden 01187, Germany}

\affiliation{$^{7}$LPTHE, CNRS and Sorbonne Université, 75252 Paris Cedex 05, France}

\affiliation{$^{*}$Correspondence: preden.roulleau@cea.fr}

\affiliation{$^{\dagger}$These authors contributed equally}

\begin{abstract}

Although individual skyrmions are topologically protected objects, their cooperative crystalline order is fragile, easily disrupted by thermal fluctuations or other external perturbations. Probing the internal dynamics of such a crystal is both compelling and challenging, as its intricate and delicate spin texture must remain stable during measurement. Here, we engineer a nanoscale graphene junction hosting a skyrmion Wigner crystal, embedded between magnon emitters and detectors. The skyrmion crystal geometry leaves a striking imprint on magnon transport: as the gate voltage is varied, near-periodic windows of sharp fluctuations in magnon count are detected across the entire sample. We develop an interpretation that this results from skyrmions being added one by one to a quasi-one-dimensional array. Each burst of the fluctuations thus corresponds to the entry of an additional skyrmion, during which the lattice stiffness reduces. The impinging magnons induce—and act as a probe of—non-equilibrium collective dynamics of the crystal. These results establish a real-space probe of topological spin textures in quantum Hall-type insulating ground states via magnon transport and open opportunities to explore correlated, topologically ordered phases in moiré and multilayer graphene systems.

\end{abstract}

\maketitle

\vspace{1em}

Quantum Hall skyrmions are quasiparticles which strikingly interweave various aspects of topology and geometry in condensed matter physics. They are topologically stable, carry topologically quantized electric charge and can form lattices of different geometries \cite{Skyrme1962,Sondhi1993,Belavin1975,Girvin1999,Brey1995}. They arise from the intimate coupling between spin and charge degrees of freedom: a smooth rotation of spins in space is accompanied by a redistribution of the electronic charge density. This in particular allows probing a crystal of skyrmions with its spatially modulated spin and charge degrees of freedom via the transport of magnons across it. This is in the spirit of Rutherford scattering, where alpha particles were used to provide information on the internal structure of matter, which turned out to consist of nuclei and their electron clouds. Here, we bring to bear modern quantum transport technology with the aim of elucidating the internal structure of topological quantum matter. 

Skyrmions, initially proposed as topological solitons in certain non-linear sigma models \cite{Skyrme1962,Belavin1975}, are smooth textures of spin, realized in a plethora of two-dimensional platforms ranging from quantum Hall ferromagnets to magnetic thin films. In the former, they were predicted to occur as charged excitations of an integer quantum Hall state at unit filling, then characterized through a non-linear sigma model description of the ensuing ferromagnetism \cite{Sondhi1993}, and soon after observed in experiment \cite{Tycko1995,Barrett1995}. In the latter, skyrmions emerge due to the Dzyaloshinskii–Moriya interaction \cite{Dzyaloshinsky1958,Moriya1960}, an exchange interaction between neighboring magnetic spins. 

Unlike the skyrmions familiar from magnetic thin films, which emerge due to the Dzyaloshinskii–Moriya interaction, in quantum Hall ferromagnets, they are stabilized by the Coulomb interaction between electrons. At the Landau-level filling factor $\nu=1$, the Coulomb energy $E_c = e^2/(4 \pi \epsilon_0 \epsilon l_B) \approx 40 \; \text{meV}$ is significantly larger than the Zeeman energy $E_Z = g \mu_B B \approx 1.5 \; \text{meV}$ at $13T$, where $B$ is the applied magnetic field, $\epsilon_0$ is the vacuum permittivity, $\epsilon \approx 5$ is the dielectric constant of graphene and $l_B$ is the magnetic length. This disparity indicates that Coulomb interactions play a dominant role in determining the ground state excitations. 

In high-mobility graphene samples subjected to a strong perpendicular magnetic field, the system enters the quantum Hall regime, where the fourfold (spin and valley) degeneracy of the $N=0$ Landau level is lifted. As a result, the first two integer filling factors, $\nu=1$ and $\nu=2$, correspond to states that share the same valley polarization but have opposite spin orientations. When the quantum Hall ferromagnet is tuned slightly away from , the upper sublevels of the $N = 0$ Landau level begin to populate, as in the case of $\nu = 1 + \delta \nu$. Including the exchange Coulomb interaction energy, it can be shown that skyrmion crystal configurations form the ground state in such systems, resulting in a Wigner crystal with non-trivial real space topology \cite{Brey1995,Yang2006}. 

Few experiments have presented evidence of a Wigner crystal of such topological objects using various techniques. In particular, heat capacity measurements \cite{Bayot1996} have revealed a low-temperature peak near the filling factor $\nu \approx 1$. Evidence for a new branch of collective spin excitations below the Zeeman gap has also been presented through strong enhancement of nuclear spin relaxation observed in two-dimensional electron gases \cite{Barrett1995} near $\nu = 1$, and also by inelastic light-scattering spectroscopy \cite{Plochocka2009} and microwave spectroscopy \cite{Kukushkin1997}. However, a real space probe which directly couples to the spin topology and the collective dynamics of the Wigner crystal ground state has been missing. The recent remarkable advent of magnon scattering experiments promises to be such a probe and has achieved some success in this regard \cite{Wei2018,zhou2020solids,Assouline2021}

The primary goal of the present work is to utilize magnon transport in a graphene quantum Hall heterojunction tailored such that it permits both magnon transmission and reflection. This provides a direct demonstration of two remarkable phenomena. First, the modification of magnon propagation arising from the nanoscale modulation of spin orientations in a skyrmion crystal—identifying the skyrmion (Wigner) crystal as a magnonic metamaterial. Second, the delicate sensitivity of the detected magnon flux to the discrete internal dynamics and structural transitions of the skyrmion crystal, witnessed through near periodic appearance of sharp noise signals: we find that individual skyrmions can be sequentially added to a quasi-one-dimensional crystalline array by tuning the electrostatic potential of the junction. This controlled addition gives rise to near-periodic, correlated fluctuations that reflect changes in the skyrmion number. The magnon signal thus provides insights into the skyrmion lattice geometry, which has been largely inaccessible so far. \\

\subsection*{Non-local magnon response and electric-field dependence}

The rich sequence of quantum Hall plateau in graphene, stemming from spin and valley degeneracies lifting in the low-lying Landau levels, makes it an ideal platform for studying collective excitations, such as magnons at filling factor $\nu = 1$ \cite{Kallin1984}. Recent experiments have successfully demonstrated magnon generation through an out-of-equilibrium occupation of edge channels. Magnon detection is usually performed by measuring absorption in the local vicinity of ohmic contacts \cite{Wei2018} or dephasing of an electronic Mach-Zehnder interferometer \cite{Assouline2021}. The threshold energy required to excite these propagating collective modes is typically given by the Zeeman energy, $E_Z = g \mu_B B$ . Magnons in quantum Hall ferromagnets are known to carry an electric-dipole moment, $\bm{p} = |e|l_B^2 \hat{\bm{z}} \times \bm{k}$, where $\bm{k}$ is the center-of-mass momentum and $\hat{\bm{z}}$ is collinear with the applied magnetic field \cite{Kallin1984,Assouline2021}. Since this electric dipole interacts with charged localized states, which often form in the bulk of the sample, magnons are expected to be scattered by charged skyrmions. This interaction provides a compelling route to investigate skyrmion dynamics and their influence on collective excitations in these systems.

The sample operates in the Quantum Hall Effect (QHE) regime (at $B = 13$ T for most of the experiments), where current flows along edge channels, as shown in Fig. 1A. It consists of a monolayer graphene flake tuned into a bipolar quantum Hall state. This configuration is achieved by using right and left bottom graphite gates, separated by $70$ nm, which allow for independent tuning of the filling factors on the left and right halves of the graphene sheet through the applied gate voltages, $V_{left}$ and $V_{right}$. In most experiments, magnon emission is driven by contact doping, with the local filling factor near the ohmic contacts being . To ensure reliable magnon emission, the electronic density of the graphene near the ohmic contacts is further tuned by an additional silicon gate positioned approximately 350 nm below the graphene flake. By applying a voltage $V_{ds}$ to the emitter contact (denoted as $E_m$), a chemical potential difference $\mu = -eV_{ds}$ is established between the inner and outer edge states, as illustrated in red and blue in the right inset of Fig. 1A. When $|eV_{ds}| > E_Z $  (with $E_Z \approx 1.5$ meV at $13$ T), spin flips can occur between the two edge states, leading to magnon emission. The emitted magnons can be detected via a non-local signal, $\frac{dV_{NL,i}}{dI}$, where  $dV_{NL,i}$ is the AC non-local voltage drop measured at the $i$-th detector contact with neighboring contacts grounded, and $dI$ the small AC current bias on top of the DC bias current $I_{ds} = V_{ds} \times e^2/h$. In the present sample, magnons are injected into a small junction with a length of $700$ nm and a width of $70$ nm, situated between the larger right and left regions. The magnons are then scattered and detected at the contacts NL1, NL2, NL3 and NL4 (see left inset of Fig. 1A).

We investigate the non-local resistance as a function of the electrostatic potential difference between the left and right sides of the junction, controlled by the two independent bottom graphite gates. The right gate is held fixed while the left gate voltage is swept, thereby tuning the potential and carrier density on the left side of the junction. Both sides are maintained on the quantum Hall ferromagnetic plateau at $\nu_L \approx \nu_R \approx 1$. In this regime, a finite in-plane electric field $E_x$ develops across the junction due to the potential difference between the two gated regions, while the magnon emission is performed remotely at an ohmic contact far from the junction.

Following the approach of Ref.\cite{Assouline2021} the absorption of a magnon at a non-local contact generates a chemical potential shift $\epsilon_i$ in the outgoing edge channel (see left inset of Fig. 1A). Similarly, absorption at the grounded contact upstream produces a shift $\epsilon_0$. The measurable non-local response is therefore $\frac{dV_{NL,i}}{dI} = (\frac{d \epsilon_0}{d\mu} - \frac{d \epsilon_1}{d\mu} ) \times h/e^2$ where $\mu = -eV_{ds}$.

Figure 1B shows $\frac{d V_{NL,i}}{dI}$  for $i = 1,2,3$ as a function of $V_{ds}$ at $\nu_L = 1.009, \; \nu_R = 1
$. A clear non-local signal appears above $eV_{ds} \approx 1.5 meV$, corresponding to the Zeeman energy $E_Z
$, marking the threshold for magnon emission. The onset of this signal demonstrates that the detected non-local response originates from spin-wave excitations (magnons) propagating through the bulk of the quantum Hall ferromagnet, rather than from charge transport along the edge channels. These bulk magnons, generated remotely at the emitter contact, travel across the sample and are absorbed at the detector, where they modulate the local chemical potential.

Figure 1C shows $\frac{d V_{NL,i}}{dI}$ as a function of the left-gate voltage, which controls both $\nu_L$ and the potential profile across the junction. The three non-local detectors display distinct behaviors, highlighting that magnon propagation depends sensitively on both sample geometry and the electric field  induced by the gate imbalance. Magnons in a quantum Hall ferromagnet carry an electric dipole moment $\bm{p} = |e|l_B^2 \hat{\bm{z}} \times \bm{k}$, and their interaction with $E_x$ depends strongly on the propagation direction. Magnons with larger transverse momentum $q_y$ experience stronger coupling to the field and therefore undergo enhanced scattering within the junction (see in Fig. E.D.4). This can be explained by an effective Lorentz force and scalar potential term arising due to the gradients of the local electrostatic potential in the junction (see Methods section ‘Influence of the Electric Field’, ‘Effects of electric field gradients in the dipole picture’ and Extended Data Fig.1). These observations demonstrate that the non-local signal above the Zeeman threshold arises from electrically tunable magnon propagation, and that the electric field across the junction acts as a controllable scattering potential whose strength depends on the relative gate voltages defining the left and right $\nu \approx 1$ regions.

\subsection*{Transition near  and emergence of skyrmion crystallization}

We next focus on magnon propagation as the left filling factor is tuned through $\nu_L = 1$. A qualitatively different behavior emerges for $\nu_L = 1 + \epsilon$, where quasiparticles are introduced into the upper spin branch of the $\nu = 2$  Landau level as represented in Fig 2A. In this regime, the added quasiparticles soften the spin stiffness of the ferromagnetic $\nu = 1$ background and favor a continuous, noncollinear spin texture. As a result, the junction hosts a skyrmion-crystal phase, where the spin rotates smoothly in space to accommodate the excess charge, as illustrated in Fig. 2B. Under these conditions we measure a sharp transition around $\sigma_{left} = 1.01 e^2/h$ and this feature is already visible in the global sweeps shown in Fig 1C where sharp fluctuations appear. To highlight its structure, we define a differential non-local resistance $\delta R_{NL2} = \frac{dV_{NL,2}}{dI} - \langle \frac{dV_{NL,2}}{dI}\rangle$  , where the average is Gaussian-smoothed to remove slow background variations. Fig 1D shows $\delta R_{NL2}$  within the sharp fluctuation region indicated by the grey box in fig 1C.  To resolve the structure of this transition in greater detail, we focus on the corresponding region in Fig. 2, D and E. These panels show a higher-resolution measurement over a narrower range of left gate voltage (highlighted in light blue in Fig. 2C) for various silicon gate voltages $V_{Si}$, which locally tune the electrostatic potential at the junction. The data reveal a series of narrow and periodic windows of strong fluctuations in the non-local signal, indicating a discrete reorganization of the underlying many-body ground state. The regions of strong fluctuations shift systematically with $V_{Si}$, indicating that the underlying instability is localized to the junction region, where the silicon gate exerts its electrostatic influence. Figure 2E shows that $\delta R_{NL2}$ exhibits a pronounced enhancement accompanied by strong periodic fluctuations in $V_{left}$. Figure E.D. 3 presents the raw data from Figures 2D and 2E without guide lines to improve visibility, together with the bias-averaged $\delta R_{NL2}$ traces for several $V_{Si}$ values, which more clearly reveals the oscillation periodicity.

In a quantum Hall ferromagnet, adding charge near $\nu = 1$ necessarily involves spin flips; however, the lowest-energy configuration of a single flipped spin is not an isolated quasiparticle but a smooth spin texture carrying unit charge—a skyrmion.  Exchange interactions lower the Coulomb energy by delocalizing the spin twist over a finite area and give rise to a topological charge density associated with each skyrmion. The characteristic skyrmion size is a few times the magnetic length $l_B = \sqrt{h/(eB)}$ ( $\approx 7$ nm at $13$ T). The lithographically defined separation between the two graphite gates is about $70$ nm, which should be regarded as an upper bound for the region in which skyrmions can form. The actual electrostatically active width is likely smaller—on the order of a few tens of nanometers ($\approx 3-10 \; l_B$). Because this dimension is comparable to the expected skyrmion diameter (typically a few $l_B$), the junction effectively acts as a one-dimensional confinement potential, favoring the formation of a linear skyrmion array along the junction.

To further clarify the origin of this confined skyrmion region, we performed self-consistent electrostatic simulations of the complete device geometry, including the graphite split gates and the silicon back gate (see Supplementary Information). The calculations show that, in the experimentally relevant $(1+\epsilon,1)$ regime, the junction develops a localized density enhancement confined to the exposed region above the gate gap, with a spatial extent of only a few magnetic lengths. As a result, the local filling factor in the junction differs from that of the surrounding bulk, creating the conditions for skyrmion crystallization. The confinement is strongest for small positive $\epsilon$, in good agreement with the narrow gate-voltage window where the enhanced magnon fluctuations are observed experimentally.

\subsection*{Global nature of fluctuations and energy scaling of the magnon noise}

To confirm that the observed fluctuations are intrinsic to the junction rather than local impurity effects, we repeat the measurement using a different magnon injector (Methods section ‘Influence of the Electric Field’ and Extended Data Fig.1). For each emitter the fluctuating response near $V_{left} = 1.36$ V persists and is observed simultaneously in all non-local detectors (Fig. 3A), indicating that the phenomenon extends across the entire device rather than being tied to a single contact (Fig. 3C). As shown in the single traces of Fig. 3B, the fluctuations are correlated across all detectors and display a striking asymmetry between the top and bottom detectors. Such correlated and asymmetric noise across multiple detectors is captured by our phenomenological model (presented below) as seen in the simulated magnon trajectories in Fig. 3D-E. We discuss the theoretical origin of such cross-detector noise in the next section.

 We further investigate the statistical properties of these fluctuations by repeating the measurement $400$ times for each drain–source bias $V_{ds}$. For a fixed gate voltage, the variance of $\frac{dV_{NL}}{dI}$ increases linearly with $V_{ds}$ once the bias exceeds the Zeeman threshold, reflecting the growing population of magnons emitted into the junction (Figs. 4A–C). The linear relation for all non-local detectors implies that the dominant scattering process occurs at the junction center, where magnons interact most strongly with the skyrmion array. This linear regime persists over a few millivolts, beyond which the variance saturates. The saturation likely arises from the gradual breakdown of the quantum Hall ferromagnetic state at large bias, where strong electric fields and enhanced Joule heating reduce spin polarization and suppress coherent magnon emission. This behavior thus delineates the crossover from a magnon-mediated non-local regime to the onset of quantum Hall breakdown. 

As magnon emission itself is a Poissonian process, one could postulate that the noise in the initial conditions of the incident magnon causes such a linear variance with $V_{ds}$ \cite{Assouline2021}. However, such stochastic noise would appear over the full gate-voltage range. In contrast, we observe pronounced fluctuations only within narrow voltage windows with noise being negligible for the majority of the observed $V_{ds}$ window. Hence, we can rule out the initial condition or partition noise as the source of such sharp, narrow and near periodic noise. In the next section we argue that the discrete process of adding a skyrmion onto an existing crystalline configuration in the junction is the most likely explanation for such narrow, near-periodic windows of noise. We argue that at these points the crystal softens and the magnons induce non-equilibrium dynamics in the crystal. Finally, a disordered “skyrmion liquid” would not exhibit such periodic sharp fluctuations upon gate tuning. Figures 2.D-E thus present direct evidence of a crystalline phase in the junction.

\subsection*{Phenomenological modeling of magnon–skyrmion interaction and noise asymmetry
}

To qualitatively capture these observations, we model the junction as a one-dimensional array of flux tubes, each representing the emergent magnetic flux associated with a skyrmion and interacting with charged magnons. The theory for a quantum Hall heterojunction containing a two-dimensional skyrmion crystal was developed in Ref.18, where it was shown that above certain threshold energies, the crystal can enhance magnon transmission owing to its modulated topological charge density profile. However, since we have a quasi-one-dimensional junction structure in this experiment, we resort to a phenomenological description of the problem inspired from the analysis  in Ref.\cite{Chakraborty2023Magnon}. We also assume a one-dimensional quadratic confining electrostatic potential along the width of the junction and a hard-wall potential beyond a certain length, along the length of the junction. The magnons feel a net Lorentz force, which has two contributions, one from the spin texture and the other from the scalar electrostatic potential modulation in the junction. The electrostatic potential also induces an effective scalar field for magnon dynamics. This framework reproduces the dipolar nature of magnons and their coupling to the skyrmion array’s topological charge density (Methods section ‘Phenomenological model involving charged particle in a flux-tube array’).  

Skyrmions carry smeared point charges, and hence, the net magnetic field resulting in the effective Lorentz force described above (Methods section ‘Dipole moment -electric field picture and non-linear sigma model: Two semiclassical limits’) is spatially modulated. Consequently, some magnons which impinge on the junction far away from the skyrmion cores pass through with minimal deflection, whereas those that are incident closer to the skyrmion core are deflected maximally. Therefore, the non-local magnon signal is detected across all detectors, we see this qualitatively in the plot of the magnon trajectories in Fig. 3D.

Since we observe significant noise and such linearly increasing variance only at very narrow and near-periodic gate voltages, we picture that such sharply enhanced noise arises due to softening of the crystal at voltages corresponding to the addition of a skyrmion to the junction. At these voltages the crystal softens, magnons passing through have an enhanced back-action, they shake the crystal. Consequently, the effective magnetic field profile fluctuates in time greatly enhancing the scattered signal of the magnons, across all detectors (see also in Fig. E.D.5). We verify this qualitative behavior in our phenomenological model in Fig. 4C, by calculating the variance of the magnon count across different detectors in the low-stiffness regime of the flux tube array. As we see, the variance across all detectors increases with increasing magnon flux. The hypothesis of non-equilibrium dynamics of the crystal at soft points induced by magnon back-action is bolstered by the period doubling of these narrow noisy windows (Figs 2.D-E), which can be explained through a structural transition from a single column to a double column.

 Moreover, in Fig. 4C we find that the simulated noise distribution exhibits a skewness, reflecting asymmetric scattering events and qualitatively matching the non-Gaussian noise statistics observed experimentally (Figs. 3B-4B). In the regime where magnons can be described as sequentially scattered particles, the asymmetry between the top and bottom detectors (NL1 and NL3) is expected to increase with the number of skyrmions (Fig. 4D).  This arises because skyrmions act as localized sources of topological charge density, deflecting magnons. As the skyrmion density increases, the junction goes closer to the uniform field limit, corresponding to maximal asymmetry. Such tunability of magnon transmission highlights the magnonic metamaterial properties of the Wigner crystal, induced by the skyrmion topology. Future experiments could track the evolution of this asymmetry with gate voltage to extract the effective size of individual skyrmions.

\subsection*{Discussion and outlook}

We have shown that a skyrmion Wigner crystal can act as a metamaterial for magnons. This has allowed us to obtain microscopic information on the geometry of the crystal prepared in a controllable fashion in a patterned heterostructure. Particularly notable is the non-equilibrium nature of our noise signal: it is strongly peaked when the crystal supports soft excitations as its geometry changes. These data naturally suggest the identification of local (addition of a Skyrmion) and global (rearrangement/change in the number of rows of the crystal) changes of the crystal geometry. The absence of the noise signal entirely when the filling is tuned to $\nu = 1$ on either side of the junction indicates that magnons propagate in a ferromagnetic background, being primarily sensitive to in-plane electric fields, in good agreement with the fact that magnons carry an electric dipole perpendicular to their momentum.

More detailed analysis of magnon transmission and reflection could yield further insights into the collective magnon-Skyrmion crystal dynamics. Other excitation mechanisms recommend themselves, e.g. using surface acoustic waves to generate phonons which in fact correspond to magnetophonons involving the spin degrees of freedom. Time-resolved profiles, as well as probing possible hysteresis on longer timescales, would further elucidate energetics and dynamics of these systems. 

Generalized Wigner crystals have received renewed experimental and theoretical interest in the recent surge of work in moiré platforms and other layered graphene samples \cite{Reddy2024,Paul2023,Guerci2025,Pan2020,Wu2019,Dong2024AHC,Dong2024QAH,Zhou2024,Desrochers2025,Xiang2025}. Our work on the nonlocal magnon transport for the skyrmion Wigner crystal in monolayer graphene opens the door for extensions to more involved platforms, e.g. those involving a different combination of internal degrees of freedom, presenting a plethora of both theoretical and experimental opportunities for such scattering problems. Further, it presents a new mesoscopic window to probe the spin structure, topology and geometry, as well as dynamics of insulating two dimensional materials in the quantum Hall regime.  Finally, future work could explore several questions with a quantum information flavour, using such magnon transport through skyrmions. For example, the quantum Hall regime of graphene (and layered extensions) has been theorized to host several spin-valley entangled phases \cite{Lian2017,Murthy2017,Stefanidis2023} with some exhibiting anisotropic ordered patterns thereof \cite{Chakraborty2024Smectic}. Future work could explore if magnon transport could act as an entanglement witness for such phases. Skyrmions, if made small enough could also realize various types of qubits \cite{Psaroudaki2021,Chakraborty2025Qubits}, and hence such magnon-skyrmion interaction, and recent proposals of magnon transduction in such systems \cite{Canright2026}, could also serve as a tool for qubit control and coupling in quantum Hall systems.

\begin{figure*}
    \centering
    \includegraphics[width=1\textwidth]{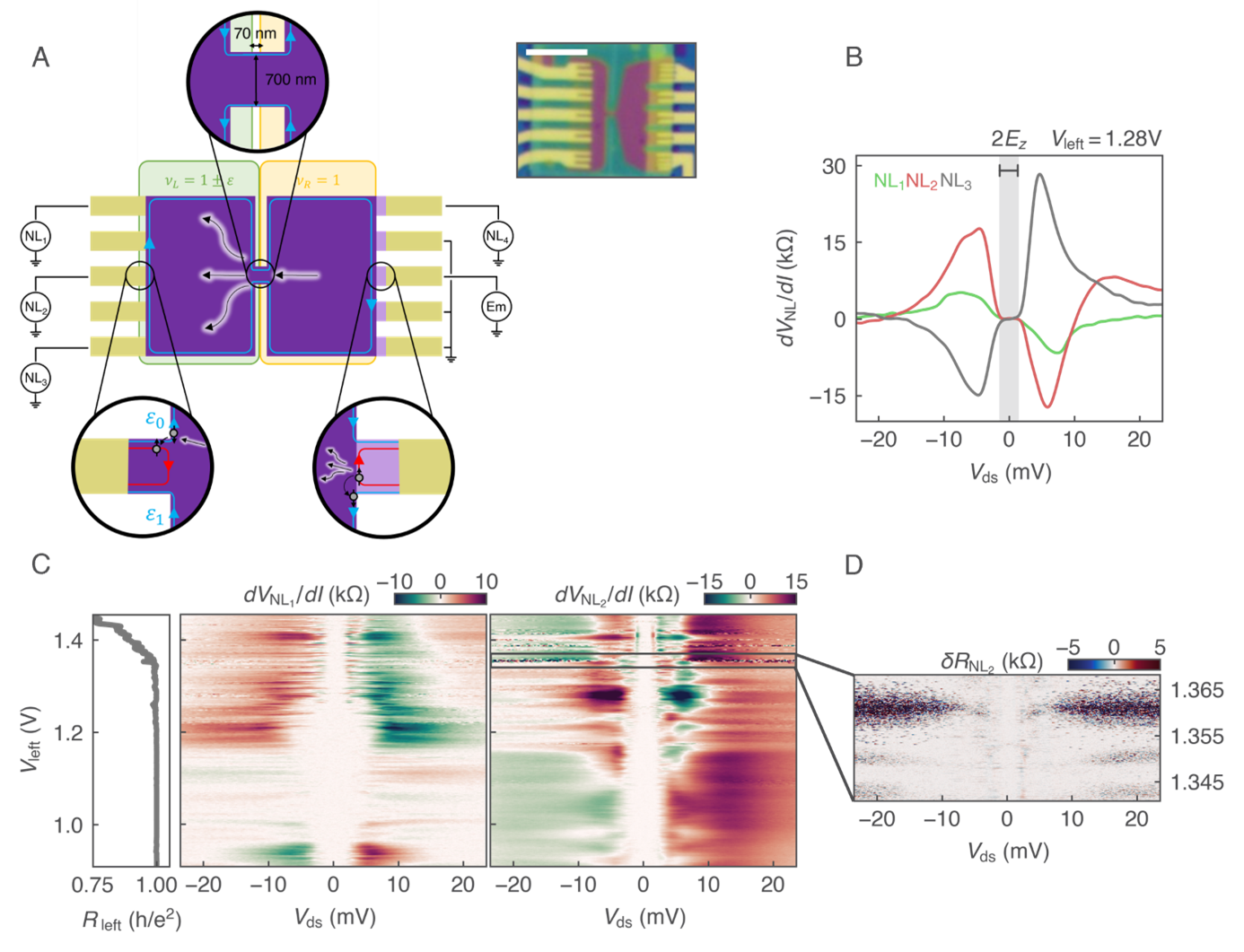}
    \caption{
    \textbf{Device schematic and magnon detection.}
    \textbf{a,}
    Schematic of the device. A monolayer graphene flake is encapsulated between hexagonal boron nitride (hBN) layers and placed on a patterned graphite back gate split into two electrically isolated sections separated by a $\sim70$\,nm gap. Independent gate voltages control the carrier density on the left and right sides of the graphene. A silicon gate beneath the right side locally increases the filling factor near the ohmic contacts to enhance magnon emission. The graphene is etched above the split gate to define a 700\,nm-wide constriction. Magnons are injected at the emitter contact, where the local filling factor is tuned to $\nu>1$, and are detected nonlocally at contacts NL1, NL2 and NL3 after scattering through the constriction.
    \textbf{b,}
    Nonlocal signals measured at detectors NL1, NL2 and NL3 as a function of DC bias at fixed gate voltage. A detectable signal appears once the applied bias exceeds the Zeeman energy ($\sim1.5$\,meV), corresponding to the onset of magnon emission.
    \textbf{c,}
    Nonlocal signals measured at detectors NL1 and NL2 as a function of the DC bias applied to the emitter and the voltage applied to the left gate. The line cut shows the corresponding two-probe resistance measured on the left side of the device while the right gate voltage is kept fixed.
    \textbf{d,}
    Sharp fluctuations, obtained after subtracting a Gaussian-smoothed background from the data shown in panel \textbf{c}, observed in detector NL2 around specific gate voltages.
    }
    \label{fig1}
\end{figure*}

\begin{figure*}
    \centering
    \includegraphics[width=1\textwidth]{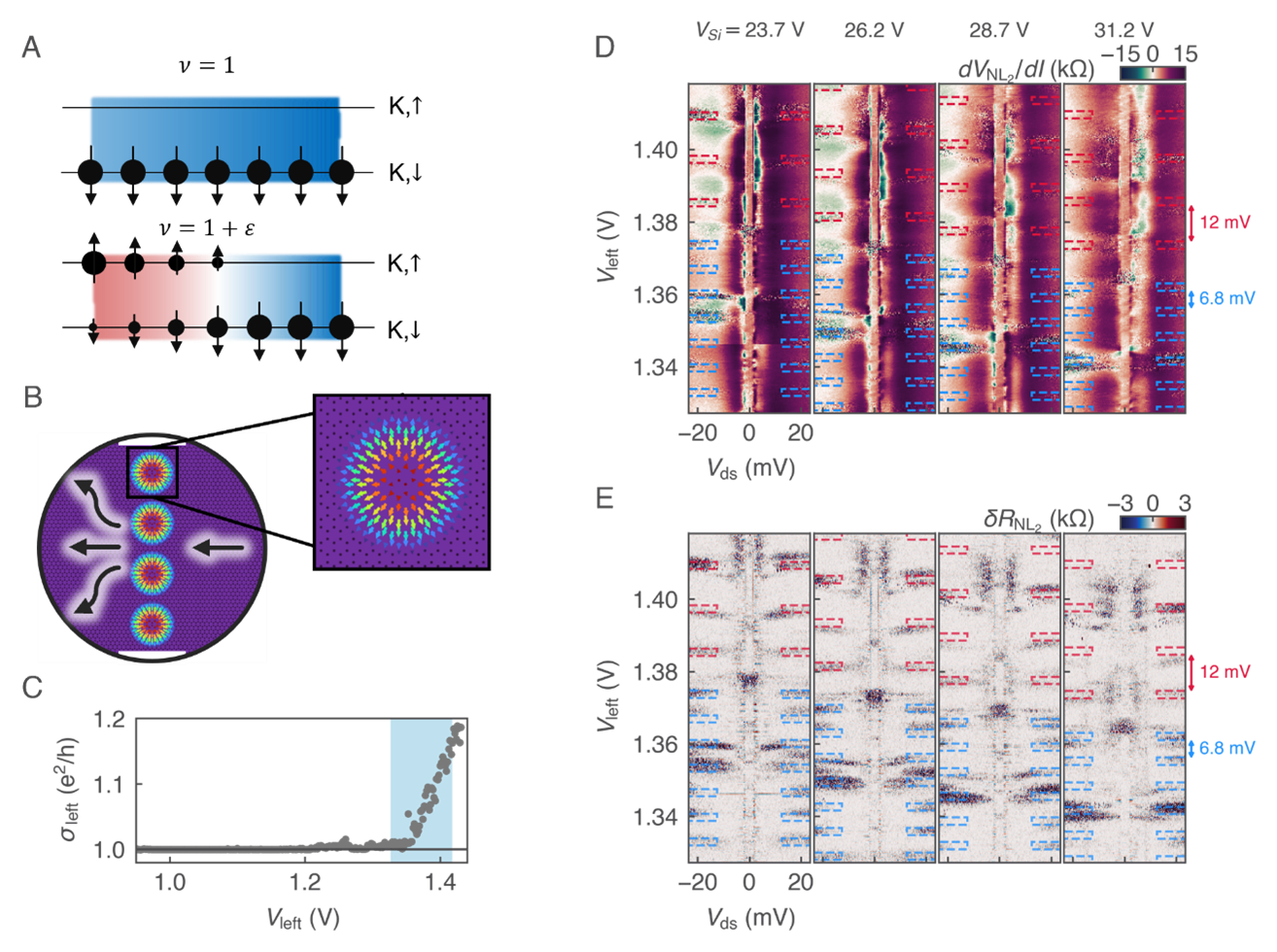}
    \caption{
    \textbf{Magnon response probing the real-space topology of a skyrmion crystal.}
    \textbf{a,b,}
    Schematics of the skyrmion crystal junction. Upon doping the lowest Landau level beyond $\nu=1$, the added charge is accommodated by topological spin textures (skyrmions) (\textbf{a}). Slightly away from $\nu=1$, these skyrmions crystallize into a topological Wigner crystal confined within the junction (\textbf{b}). Magnons emitted from a single injector (Em2) are scattered by the crystal and detected through the nonlocal response.
    \textbf{c,}
    Hall conductance as a function of the left-gate voltage. The light-blue region, slightly away from $\nu=1$, marks the gate-voltage window where sharp fluctuations in the nonlocal signal are observed.
    \textbf{d,e,}
    Differential nonlocal signal measured at detector NL2 (\textbf{d}) together with the corresponding background-subtracted signal (\textbf{e}) as a function of the DC bias and left-gate voltage, for several silicon-gate voltages. Near-periodic noisy regions emerge as individual skyrmions are added sequentially to the crystal confined within the junction. Beyond a critical gate voltage, the oscillation period doubles, indicating a structural transition from a one-dimensional skyrmion chain to a two-row crystal. The enhanced temporal fluctuations arise from magnon-induced dynamics of the skyrmion crystal at gate voltages where the lattice becomes soft.
    }
    \label{fig2}
\end{figure*}

\begin{figure*}
    \centering
    \includegraphics[width=1\textwidth]{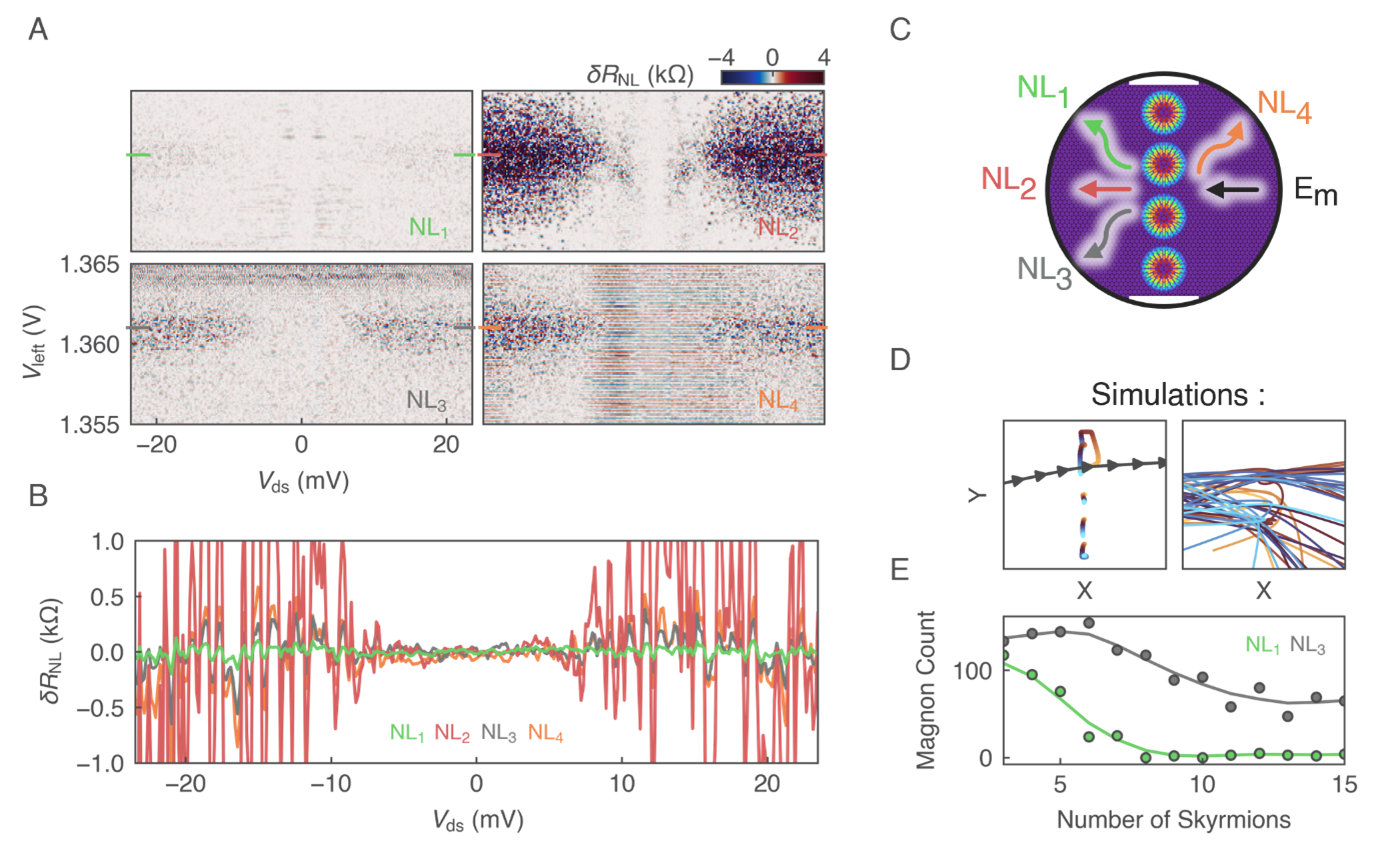}
    \caption{
    \textbf{Spatially correlated magnon fluctuations induced by a skyrmion crystal.}
    \textbf{a,}
    Background-subtracted nonlocal resistance measured at detectors NL1, NL2, NL3 and NL4 as a function of drain--source voltage.
    \textbf{b,}
    Fluctuations in the nonlocal signal measured at fixed gate voltage as a function of drain--source voltage. The fluctuations are strongly correlated across all detectors, demonstrating a common microscopic origin.
    \textbf{c,}
    Schematic of magnon propagation through the junction. Transmitted magnons are detected by nonlocal detectors NL1, NL2 and NL3 on the left side of the junction, whereas reflected magnons are detected by NL4.
    \textbf{d,e,}
    Phenomenological model of magnon scattering by a skyrmion crystal, represented as a flux-tube array.
    \textbf{d,}
    Left: trajectories of five skyrmions perturbed by the passage of a single magnon (grey). The skyrmions, initially confined to a narrow strip, evolve in time from blue to orange. Right: trajectories of forty magnons carrying an electric dipole moment after scattering from the skyrmion crystal.
    \textbf{e,}
    Simulated asymmetry of the scattered magnon signal, quantified by the imbalance between the top (NL3) and bottom (NL1) detectors, as a function of the number of skyrmions in the junction. The asymmetry increases with skyrmion number, reflecting the growing influence of the crystal on magnon trajectories. See the main text and Methods for details.
    }
    \label{fig3}
\end{figure*}

\begin{figure*}
    \centering
    \includegraphics[width=1\textwidth]{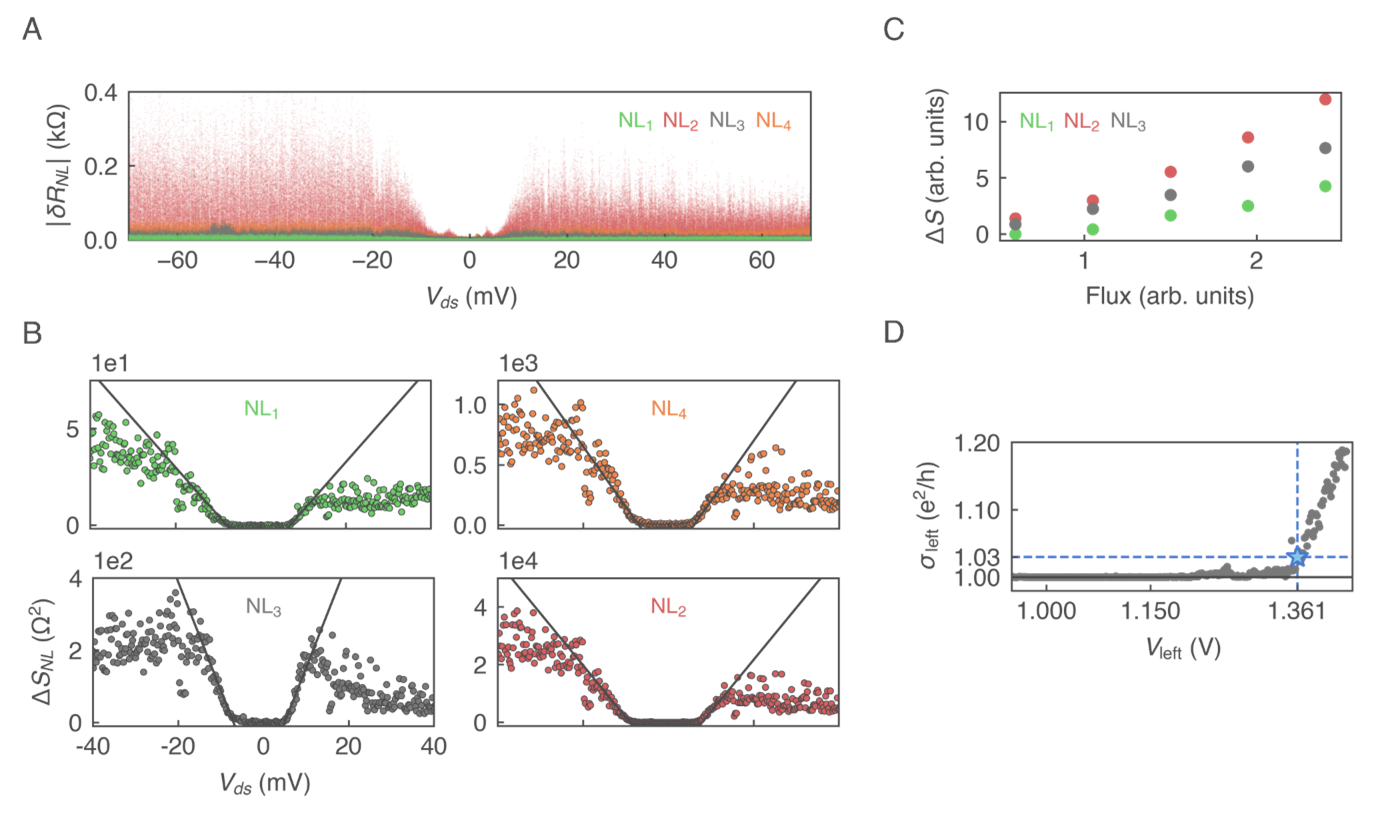}
    \caption{
    \textbf{Noise signatures of magnon--skyrmion interactions in a low-stiffness crystal.}
    \textbf{a,}
    Background-subtracted nonlocal magnon signal measured simultaneously at detectors NL1, NL2, NL3 and NL4, averaged over 400 repetitions at each bias point. Sharp fluctuations emerge within narrow gate-voltage windows and are attributed to magnon-induced non-equilibrium dynamics of the skyrmion crystal. As magnons perturb the soft crystal, the resulting fluctuations in the effective scattering potential generate correlated noise across all detectors.
    \textbf{b,}
    Variance of the nonlocal signal as a function of drain--source voltage. Above the magnon-emission threshold, the variance increases linearly with bias before saturating at higher voltages.
    \textbf{c,}
    Phenomenological model showing the variance of the scattered magnon signal as a function of magnon flux for a low-stiffness skyrmion crystal (flux-tube array). The calculated variance increases linearly with magnon flux for all three detectors, in agreement with the experimental observations.
    \textbf{d,}
    Hall conductance as a function of the left-gate voltage. The blue star marks the gate voltage at which the measurements shown in panels \textbf{a} and \textbf{b} were performed.
    }
    \label{fig4}
\end{figure*}

\section*{Acknowledgements}

We thank Allan H. MacDonald for insightful discussions on the theoretical modeling. This work was supported by the European Research Council (ERC Starting Grant COHEGRAPH, No.~679531), the Agence Nationale de la Recherche (ANR EQUBITFLY), the Horizon Europe EIC Pathfinder project FLATS (HORIZON-EIC-2022-PATHFINDEROPEN), and the Horizon Europe EIC Pathfinder project ELEQUANT (HORIZON-EIC-2024-PATHFINDEROPEN).

\section*{Author contributions}

P.R. conceived and supervised the project. R.A. and P.R. performed the experiments. Q.B. and M.K. fabricated the device. A.L.-S. and L.S. carried out the electrostatic simulations under the supervision of C.G. N.C., R.M. and B.D. developed the theoretical model with input from P.R. R.A., M.K., Q.B., H.C., L.P. and P.R. analysed the data. K.W. and T.T. provided the hexagonal boron nitride crystals. R.A., N.C., B.D. and P.R. wrote the manuscript with input from all authors.

\section*{Competing interests}

The authors declare no competing interests.

\section*{Data availability}

The data and analysis code supporting the findings of this study are available on Zenodo (Ref.\cite{Ayache2025}).
\onecolumngrid

\clearpage

\clearpage
\clearpage
\begin{center}
    {\LARGE\bfseries Materials and methods}
\end{center}

\input{materials_and_methods}

\clearpage

\begin{center}
    {\LARGE\bfseries Supplementary Information}
\end{center}

\input{SI}

\clearpage
\begin{center}
    {\LARGE\bfseries Extended Data}
\end{center}

\input{Extended_data}
\clearpage
\bibliography{references,supp_mag}

\end{document}

%% file: materials_and_methods.tex
\section{Device Fabrication and geometry}
High-quality hBN/graphene/hBN/graphite heterostructures were prepared via mechanical exfoliation followed by a dry transfer method. Monolayer graphene and few-layer graphite were obtained by exfoliating natural graphite (NGS GmbH) onto pre-cleaned Si/SiO$_2$ substrates with a 285 nm oxide layer. In a similar manner, hexagonal boron nitride (hBN) flakes were exfoliated from bulk hBN crystals (NIMS, Japan) onto separate cleaned Si/SiO$_2$ substrates. The exfoliated flakes were initially characterized using optical microscopy. To assess flake uniformity and imperfections, dark-field and differential interference contrast imaging were employed. Flakes exhibiting uniform contrast and absence of visible cracks were selected for further inspection. These selected flakes were subsequently characterized using non-contact mode atomic force microscopy (AFM) to verify surface quality, ensure the absence of microcracks or surface contamination, and confirm thickness uniformity across the flake. A suitable graphite flake was selected and transferred onto a pre-patterned p-doped Si/SiO$_2$ substrate using a polypropylene carbonate (PC)/polydimethylsiloxane (PDMS) stamp. The graphite flake was then patterned into two independent local bottom gates denoted as left and right gate using electron beam lithography (EBL) followed by oxygen plasma etching. The two gate regions were separated by a narrow constriction slit approximately 70 nm wide, as shown in Fig.~\ref{fig:S1}a. Following the patterning, the graphite flake was cleaned in hot acetone for several hours to remove resist residues, and then annealed under vacuum at 350$^\circ$C to eliminate surface contaminants. The patterned structure was again examined using non-contact AFM to ensure the uniformity of the flake. Next, a van der Waals heterostructure consisting of monolayer graphene encapsulated between two hexagonal boron nitride (hBN) layers ($\sim$37 nm thick on top and $\sim$60 nm at the bottom) was assembled and aligned precisely over the pre-patterned graphite gates using the PC/PDMS stamp. The stack was then cleaned, followed by vacuum annealing at 350$^\circ$C to remove polymer residues and improve interfacial cleanliness. The fabricated stack was subsequently patterned using EBL and reactive ion etching (RIE) with CHF$_3$/O$_2$ gas to shape the graphene. A second EBL step was performed to design one-dimensional edge contacts, followed by another CHF$_3$/O$_2$ RIE step to selectively etch and expose the graphene edges. The contacts were made in a finger-like geometry to increase the contact area and reduce the contact resistance. Cr/Au (10/70 nm) was deposited by electron beam evaporation at an angle using a tilted stage to ensure sidewall coverage. The graphene channel was subsequently defined through an additional EBL step and etched using the same RIE recipe. An optical image of the completed device is presented in Fig.~S.1b and a schematic of the heterostructure stack is shown in Fig.~S.1c.

\section{Influence of the Electric Field }

To examine how the in-plane electric field affects magnon propagation across the junction, we emit magnons from three contacts ($E_{\mathrm{m,1}}$, $E_{\mathrm{m,2}}$, and $E_{\mathrm{m,3}}$) arranged as shown in Fig.~E.D.1a and detect the resulting non-local response at NL2. For clarity the schematic depicts all emitters simultaneously, but in the experiment each dataset was obtained independently by activating only a single emitter at a time. The strongest non-local signal is observed for the lateral emitters $E_{\mathrm{m,1}}$ and $E_{\mathrm{m,3}}$, whose emission directions intersect the junction at oblique angles. In this geometry, the magnons acquire a larger transverse momentum component $q_y$, which increases the effective dipolar moment associated with their precession and strengthens the coupling between their electric dipole $\mathbf{p}$ and the in-plane electric field $E_x$ across the junction. Because the dipole--field interaction scales as $\mathbf{p}\cdot\mathbf{E}$, magnons with finite $q_y$ experience a stronger field-induced torque than those approaching the junction nearly normal to it. This enhanced coupling promotes elastic mode mixing and trajectory bending within the junction region, effectively redirecting magnons and increasing their probability of being absorbed or detected at NL2. In addition, the field can transfer spectral weight from long-lived propagating modes into more strongly damped modes, further reducing the transmitted magnon population.

Additional attenuation may originate from the electrostatic disorder, where localized charge puddles create spatially varying electric fields and modifications of the local magnetic environment. Although magnons are charge-neutral, their dipolar nature renders them sensitive to these gradients, leading to additional scattering, deflection, or partial absorption before they reach NL2.

\section{Magnon propagation and Signal Conservation at $\nu_L = 1- \varepsilon$}
 
As a complement to the measurements performed at $\nu_L = 1 +\varepsilon$, which correspond to most of the data presented in the main text, we also investigate magnon propagation slightly below the integer filling, at $\nu_L = 1- \varepsilon$. To probe this regime, we excite the central emitter $E_{\mathrm{m,2}}$ and measure the resulting non-local signals at all three detectors, NL1, NL2, and NL3 (Fig.~E.D. 2a). The measurements are performed while simultaneously recording the Hall conductivity $\sigma_{xy}$ of the right-gated region (Fig.~E.D. 2b), providing a reference for the local electronic state.

The non-local signals at the three detectors show a remarkable feature: although individual detector amplitudes vary slightly due to geometric factors and small residual disorder, the total integrated signal summed over NL1, NL2, and NL3 remains approximately constant (Fig.~E.D. 2c). This conservation implies that the magnons emitted from $E_{\mathrm{m,2}}$ propagate through the bulk without significant loss or absorption, consistent with a high-quality, incompressible bulk state.

From this measurement perspective, we do not observe any distinctive signatures of a skyrmion crystal in the bulk: the magnon signal distribution remains smooth and conserved, without strong spatial modulation or anomalous enhancement that might indicate a skyrmion lattice.

\section{Dipole moment -electric field picture and non-linear sigma model: Two semiclassical limits
}

\subsection{Dipole picture}

This description first arose in the early paper by Kallin and Halperin~\cite{Kallin_84} and it is based on the
following Hamiltonian:
\begin{equation}
    H_{\mathrm{eff}}(\bm{r},\bm{q}) = \sqrt{\frac{\pi}{32}} \dfrac{e^2}{\epsilon l_B} (ql_B)^2 + V_{\mathrm{ext}} (\bm{r} + \frac{l_B^2}{2}\bm{z} \times \bm{q}) - V_{\mathrm{ext}}(\bm{r}-\frac{l_B^2}{2}\bm{z} \times \bm{q})
    \label{eq:H_eff}
\end{equation}
where the first term is the standard one derived in \cite{Kallin_84} for a magnon on top of a ferromagnetic background, and the second two arise in the presence of an external electrostatic potential, induced for example by gating the 2D electron gas.
In this equation, $\bm{r}$ denotes the center of mass position of the dipole formed by a spin-carrying particle-hole pair,
and $q$ is the conjugate wave-vector. A key insight of~\cite{Kallin_84} is that the relative coordinate of the dipole 
$\bm{s}=\bm{r}_{\mathrm{part}}-\bm{r}_{\mathrm{hole}}$ is directly related to the momentum $\bm{q}$ according
to $\bm{q}=l_B^2 \bm{z} \times \bm{q}$. Microscopically, this arises because all
single particle orbital degrees of freedom are constrained to belong to a single Landau level. Because guiding center
coordinates for a single electron behave as a pair of canonically conjugate variables, the classical phase-space 
associated to a particle-hole pair is four-dimensional, and can be organized in terms of the $\bm{r},\bm{q}$ pair.
The precise status of $H_{\mathrm{eff}}(\bm{r},\bm{q})$ is as the Weyl symbol of the Hamiltonian operator $\hat{H}$ acting on the
subspace of electronic Fock space corresponding to a single particle-hole pair. If $\Psi(\bm{r})$ denotes the
wave-function, then:
\begin{equation}
    (\hat{H}\Psi)(\bm{r})=\int\frac{d^2 \bm{r'} d^2\bm{q}}{(2\pi)^2}H_{\mathrm{eff}}(\frac{\bm{r}+\bm{r'}}{2},\bm{q})\,
    e^{i \bm{q.(r-r')}}\, \Psi(\bm{r'})
\end{equation}

In this work, we are not attempting to solve the associated time-dependent Schr\"odinger equation for such dipole wave-functions,
but limit ourselves to semi-classical equations of motion for localized wave-packets. This is valid in the short wave-length limit
$1 \ll qd_{\mathrm{ext}}$, where $d_{\mathrm{ext}}$ stands for the characteristic length of the spatial variations of $V_{\mathrm{ext}}$.
The equations of motion for the phase-space location of wave-packets take the usual Hamiltonian form:
$\hbar \frac{d \bm{r}}{dt}=\bm{\nabla_{q}}H_{\mathrm{eff}}$, $\hbar \frac{d\bm{q}}{dt}=-\bm{\nabla_{r}}H_{\mathrm{eff}}$.
These equations become much simpler in the limit when the separation $s$ between the particle and the hole is small
compared to $d_{\mathrm{ext}}$, so that we can replace 
$V_{\mathrm{ext}}(\bm{r}_{\mathrm{part}})-V_{\mathrm{ext}}(\bm{r}_{\mathrm{hole}})$ by 
$l_B^2 \bm{(z \times q).\nabla}V_{\mathrm{ext}}(\bm{r})$. This requires the following upper bound on the momentum:
$ql_B^2 \ll d_{\mathrm{ext}}$. In order for this upper bound to be compatible with the previous lower bound, we require
that $l_B \ll d_{\mathrm{ext}}$, which we assume to be valid throughout this work.

\subsection{Non-linear sigma model picture}

The dipole picture has been firmly established for magnon propagation through a ferromagnetic background.
In the presence of skyrmions, it is probably easier to start from a description based on the non-linear
sigma model~\cite{Sondhi_13,Moon_95}. It is based on the Lagrangian:
\begin{equation}
    L=\hbar \int \frac{d^2 \bm{r}}{4 \pi l_B^2}\,\bm{\mathcal{A}}(\bm{n})\bm{.}\partial_t \bm{n} - H,
\end{equation}
where $\bm{\nabla\times\mathcal{A}}(\bm{n})=\bm{n}$, and the energy functional $H$ is given by:
\begin{equation}
    H = \frac{\rho_s}{2} \int (\bm{\nabla n})^2 + \frac{e^2}{2\epsilon} \int d^2 \bm{r} d^2 \bm{r'}\,\dfrac{Q(\bm{r}) Q(\bm{r'})}{|\bm{r}-\bm{r'}|} + \int d^2 \bm{r}\,V_{\mathrm{ext}}(\bm{r})\, Q(\bm{r}) 
\end{equation}
In this expression, the stiffness $\rho_s = e^2/16\sqrt{2\pi}\epsilon l_B$, and
the topological charge density is given by:
\begin{equation}
 Q(\textbf{r}) = \frac{1}{4\pi} \textbf{n} \cdot (\partial_x \textbf{n} \times \partial_y \textbf{n}) \ .
 \label{topceq}
\end{equation}
The resulting equations of motion take the usual Landau-Lifschitz form:
\begin{equation}
   \frac{\partial \bm{n}} {\partial t} = \frac{4\pi l_B^2}{\hbar}\,\bm{z \times \nabla_{n(r)}}H
\end{equation}

In the case of a ferromagnetic background, the connection between the non-linear sigma model
description and the dipole picture can be understood as follows. Assuming that spins point everywhere
close to the north pole, the spin texture can be described via the small planar components
$n_x(\bm{r}),n_y(\bm{r}) \ll1$. In a quantum description, this corresponds to a local
spin-1/2 wave-function $(\Psi_{\uparrow}(\bm{r}),\Psi_{\downarrow}(\bm{r}))\simeq (1,(n_x(\bm{r})+i n_y(\bm{r}))/2)$, neglecting second order terms in $n_x(\bm{r})$ and $n_y(\bm{r})$. In this situation,
the dipole wave-function $\Psi(\bm{r})$ can be identified with $\Psi_{\downarrow}(\bm{r})$, so
the time dependent Schr\"odinger equation for $\Psi(\bm{r})$ translates into the linearized
Landau-Lifschitz equation of motion for the in plane spin deviations $(n_x(\bm{r}),n_y(\bm{r}))$.
It is instructive to check this statement by considering the effect of an external potential
$V_{\mathrm{ext}}(\bm{r})$. The linearized equations of motion can be obtained by considering the
second order variation of $H$ around an equilibrium configuration. When the latter is the ferromagnetic
state along the north pole, we get, setting $\Psi(\bm{r}) =2 \Psi_{\downarrow}(\bm{r})$: 
\begin{equation}
    H = \frac{\rho_s}{2} \int |\bm{\nabla}\Psi|^2  + \frac{i}{8 \pi}\int V_{\mathrm{ext}}\,(\partial_{x}\Psi \,\partial_{y}\bar{\Psi}
    - \partial_{y}\Psi \,\partial_{x}\bar{\Psi})
    \label{linearized_energy_funct}
\end{equation}
This leads to:
\begin{equation}
    i\hbar \frac{\partial \Psi}{\partial t}= - \sqrt{\frac{\pi}{32}} \dfrac{e^2}{\epsilon l_B} 
    (l_B \bm{\nabla})^2 \Psi + \frac{l_B^2}{2i}\, \bm{z.}(\bm{\nabla}\Psi \bm{\times\nabla}V_{\mathrm{ext}})(\bm{r})
\end{equation}
Setting $\hat{q}=-i\bm{\nabla}$, and in the long wave-length limit ($ql_B^2 \ll d_{\mathrm{ext}}$),
we recover the Schr\"odinger equation associated to the Hamiltonian operator~(\ref{eq:H_eff}) acting on the dipole wave-function $\Psi(\bm{r})$.

In the presence of a slowly varying spin texture, the magnon dynamics is modified due to the presence
of a spin Berry phase induced by spatial gradients of the spin orientation. 
This manifests as an additional orbital magnetic field, equal to the topological charge density $4\pi\,Q(\bm{r})$ acting on a magnon~\cite{chakrabortymagnon}.
As usual, we take this effect into account by substituting the covariant derivative
$-i\bm{\nabla}-\bm{A}$ to the original $-i\bm{\nabla}$ operator, where $\bm{\nabla \times A}=4\pi\,Q$.
Because skyrmions are charged objects, they generate an additional electrostatic potential,
therefore the external potential has to be replaced in the magnon equation of motion by the local potential:
\begin{equation}
   V_{\mathrm{loc}}(\bm{r})=V_{\mathrm{ext}}(\bm{r})+
   \frac{e^2}{\epsilon} \int d^2 \bm{r'} \,\dfrac{ Q(\bm{r'})}{|\bm{r}-\bm{r'}|}
\end{equation}
In the following, it will be convenient to use rescaled space-time coordinates,
where the position is expressed in units of $l_B$ and $\bm{p}$ stands for $l_B \bm{q}$. 
Since we are working in the geometrical optics approximation for the magnons, we describe them by a
wave-packet of the form:
\begin{equation}
    \Psi(\bm{r})=\theta\, e^{-\frac{(\bm{r}-\bm{r}_m)^2}{2a^2}}\,e^{i(\bm{p.}(\bm{r}-\bm{r}_m)+\varphi)}
\end{equation}
where $\bm{r}_m$ denotes the magnon position and $\bm{p}$ its momentum, $\theta$ is the amplitude of the
spin deviation away from the average magnetization vector, and $\varphi$ a global phase.
The spatial scale $a$ is chosen to be large compared to $p^{-1}$ and small compared to the
characteristic scales associated to the variations of the local potential.
Inserting this Ansatz in the above energy functional~(\ref{linearized_energy_funct}), adapted
to the case of slowly varying texture as explained above, gives:
\begin{equation}
   (\theta^2 \pi a^2)^{-1} H(\bm{r}_m,\bm{p},\theta)=\frac{\rho_s}{2}\left((\bm{p}-\bm{A})^2-4\pi\,Q_0(\bm{r})\right)
    + \frac{\bm{\hat{z}}}{8\pi}\bm{.}((\bm{p}-\bm{A})\bm{\times \nabla}V_{\mathrm{loc}})-\frac{1}{2}\,
    V_{\mathrm{loc}}(\bm{r})\,Q_0(\bm{r})
    \label{eq:energy_wavepacket}
\end{equation}
So far, the amplitude of the spin deviation $\theta$ and the size $a$ of the wave-packet are undetermined.
To lift this indeterminacy, we notice that the total magnetization in the system is given by:
\begin{equation}
    \bm{M}=\int \frac{d^2 \bm{r}}{2 \pi} \, \bm{n}(\bm{r})
\end{equation}
We now impose that, when applied on a ferromagnetic background, introducing a magnon wave-packet produces the same
total magnetization as flipping one local spin into its opposite. This leads to:
\begin{equation}
    2 = \int \frac{d^2 \bm{r}}{2 \pi} \, (1-\sqrt{1-|\Psi(\bm{r})^2|})
\end{equation}
Assuming that $\theta$ is small and performing the gaussian integral gives:
\begin{equation}
\theta^2 \pi a^2 = 8 \pi   
\end{equation}
Inserting this value in the above expression~(\ref{eq:energy_wavepacket}) gives the following magnon Hamiltonian
\begin{equation}
H_{\mathrm{mag}} = \frac{\omega_{\mathrm{int}}}{2}\bigg(\big(p_x-A_x)^2 + \big(p_y-A_y)^2 \bigg) + (p_x-A_x)\, \partial_y V_{\mathrm{loc}} - (p_y-A_y)\,\partial_x V_{\mathrm{loc}}  
\label{eq:H_magnon}
\end{equation}
where $\omega_{\mathrm{int}}=\sqrt{\pi/8}\,e^2/(\hbar \epsilon l_B)$. To simplify the notation, the magnon position is denoted here by $\bm{r}$ instead of $\bm{r}_m$.
Note that we have neglected the potential terms in~(\ref{eq:energy_wavepacket}) since the most interesting physical effects on magnon propagation are due to the presence of the effective orbital magnetic field $\bm{A}$ ~\cite{chakrabortymagnon} and of a dipole moment proportional to the linear momentum and orthogonal to it~\cite{Kallin_84}. 
The corresponding Hamiltonian equations of motion read:
\begin{equation}
     \frac{d\bm{r}}{dt} = \bm{\nabla_p} H_{\mathrm{mag}},\,\,\,\frac{d\bm{p}}{dt} = - \bm{\nabla_r} H_{\mathrm{mag}}
     \label{Hamilton_eq_magnons}
\end{equation}

From the above Hamiltonian, one can apply a Legendre transform to obtain the Lagrangian: 
\begin{equation}
    L_{\mathrm{mag}} = \frac{1}{2\omega_{\mathrm{int}}} \left((\dot{x} - v_x)^2 + (\dot{y} -v_y)^2 \right) + A_x \dot{x} + A_y \dot{y},
\end{equation}
where we have introduced the drift velocity $(v_x,v_y)$ defined by 
$v_x = \partial_y V_{\mathrm{loc}}$ and $v_y = -\partial_x V_{\mathrm{loc}}$. 
The equations of motion take the form:
\begin{equation}
\begin{aligned}
    \ddot{x} &= \omega_{\mathrm{int}}\left(\partial_x A'_y - \partial_y A'_x \right) \dot{y} + \frac{1}{2}\partial_x \big(v_x^2 + v_y^2)\\
    \ddot{y} &= - \omega_{\mathrm{int}}\left(\partial_x A'_y - \partial_y A'_x \right) \dot{x} + \frac{1}{2}\partial_y \big(v_x^2 + v_y^2)
\end{aligned}
    \label{mageqm}
\end{equation}
where $A'_x = A_x - v_x/\omega_{\mathrm{int}}$ and $A'_y = A_y - v_y/\omega_{\mathrm{int}}$. 
We see that there are two Lorentz force contributions, one arising from the twist of the spin and the corresponding Berry phase, and the other due to the drift velocity induced by the smooth variation of the external potential. 
The latter also gives rise to an effective scalar potential.

So far, we have discussed magnon dynamics around a fixed texture. However, in the presence of a
finite region containing a Skyrmion crystal, 
we have two types of low energy excitations: unbound ones, that may be injected far away from the Skyrmion crystal
and may propagate to arbitrary distances, and internal breathing modes of the crystal. For simplicity, we shall
consider only the breathing modes. These can be parametrized by $N_S$ Skyrmion positions
$\bm{R}_1,...,\bm{R}_{N_S}$, so that the above Hamiltonian dynamics associated to one magnon has to be 
extended to incorporate these magneto-phonon degrees of freedom. Concretely, this implies that the reference
spin texture should now be regarded as a function $\bm{n}(\bm{r};\bm{R}_1,...\bm{R}_{N_S})$ of $N_S +1$ positions.
In an adiabatic picture, we can in principle minimize the non-linear sigma model energy functional in the presence
of the external potential $V_{\mathrm{ext}}$ with the constraint that $N_S$ Skyrmions are centered at $\bm{R}_1,...,\bm{R}_{N_S}$, so that we get an energy function $V_{\mathrm{el}}(\bm{R}_1,...,\bm{R}_{N_S})$.
For the sake of simplicity, we have not attempted to compute this Skyrmion generalized elastic energy from
the full non-linear sigma model. We postulate a simple phenomenological form, which is based on the fact
that each Skyrmions carries a fundamental electric charge $e$, interacting with the laws of electrostatics.
Typically, we set:
\begin{equation}
    V_{\mathrm{el}}(\bm{R_1},...,\bm{R_{N_S}})= \tau \sum_{i=1}^{N_S} V_{\mathrm{ext}}(\bm{R}_i) + 
    \frac{e^2}{2\epsilon l_B}\sum_{<i,j>} \frac{1}{(|\bm{R}_i - \bm{R}_j|^2 + 1)^{1/2}}
    \label{eq:V_elastic}
\end{equation}
As above, positions are expressed in units of the magnetic length $l_B$. 
Here, $\tau=1$ (resp. $\tau=-1$) for negatively (resp. positively) charged Skyrmions.
The inter-Skyrmion Coulomb potential
is expected to be regular at small separation, which motivates the above modification 
of the bare Coulomb potential at short distances. To simplify the numerical calculations, we
will also keep only the Coulomb repulsion between nearest neighbor Skyrmions.
The magnon dynamics in the presence of $N_S$ Skyrmions is assumed to take the same form as
before, but in the magnon Hamiltonian~(\ref{eq:H_magnon}), both the effective gauge field $\bm{A}$
and the local potential $V_{\mathrm{loc}}$ depend on the Skyrmion positions. Again, we do not attempt
to extract them from a full calculation, but postulate the phenomenological forms:
\begin{equation}
 \bm{\nabla \times A}(\bm{r})   = \sum_{i=1}^{N_S} \frac{4a^2}{(|\bm{r} - \bm{R_i}|^2 + a^2)^{2}}
 \label{vecphenom}
\end{equation}
\begin{equation}
   V_{\mathrm{loc}}(\bm{r}) = V_{\mathrm{ext}}(\bm{r}) + \tau \frac{e^2}{\epsilon l_B} \sum_{i=1}^{N_S}
   \frac{1}{(|\bm{r} - \bm{R}_i|^2 + b^2)^{1/2}}
   \label{vlocphenom}
\end{equation}
Here $a$ and $b$ are positive length scales of the order of the magnetic length. For a qualitative phenomenological modelling of the electrostatic potential of the junction we shall assume that $V_{ext}$ comprises a hard-wall along the $y$-axis potential beyond a certain junction length, and a quadratic potential along the $x$-axis, which flattens out to a constant beyond the width of the junction. We present the details and results of such a phenomenological approach in the next section.

Equations of motion for Skyrmion positions can be derived by restricting the symplectic form generating
Hamilton's equations of motion from the infinite-dimensional family of smooth maps $\bm{n}(\bm{r})$ to
the finite sub-manifold of low energy configurations $\bm{n}(\bm{r};\bm{R}_1,...\bm{R}_{N_S})$ parametrized by the Skyrmion positions. The symplectic 
form $\Omega$ is defined as follows: for two infinitesimal deformations $\bm{\xi}_{1}(\bm{r})$, $\bm{\xi}_{2}(\bm{r})$
around $\bm{n}(\bm{r})$ (so $\bm{n}(\bm{r})\bm{.}\bm{\xi}_{i}(\bm{r})=0$, $i=1,2$),
\begin{equation}
\Omega_{\{\bm{n}\}}(\bm{\xi}_{1},\bm{\xi}_{2})= \int \frac{d^2 \bm{r}}{4 \pi l_B^2} \,
\bm{n} \bm{\cdot} (\partial_x \bm{\xi}_{1} \bm{\times} \partial_y \bm{\xi}_2)
\end{equation}
Let us now consider two infinitesimal variations $\delta \bm{R}_{i}^{(1)}$ and $\delta \bm{R}_{i}^{(2)}$
of the Skyrmion positions. This gives rise to $\bm{\xi}_{1}(\bm{r})$ and $\bm{\xi}_{2}(\bm{r})$
according to:
\begin{equation}
    \bm{\xi}_{1,2}(\bm{r})=\sum_{i=1}^{N_S}\sum_{a=x,y}\frac{\partial \bm{n}}{\partial \bm{R}_{i}^{a}}\, \delta \bm{R}_{i}^{a,(1,2)}
\end{equation}
Substituting in the above expression for the symplectic form, we get a complicated expression, that we may
approximate assuming that Skyrmions do not overlap much. As a first consequence, we can neglect off-diagonal
terms of the form $\delta \bm{R}_{i}^{a,(1)}\,\delta \bm{R}_{j}^{b,(2)}$ with $i \neq j$.
While evaluating the $\bm{r}$ integral, the prefactor of the 
$\delta \bm{R}_{i}^{x,(1)}\,\delta \bm{R}_{i}^{y,(2)}$ term is dominated by the region where 
$\bm{r}$ is closer to $\bm{R}_{i}$ than to $\bm{R}_{j}$ for $j \neq i$. In this region, we have 
an approximate invariance for $\bm{n}(\bm{r};\bm{R}_1,...\bm{R}_{N_S})$ under infinitesimal translations
that are acting only on $\bm{r}$ and $\bm{R}_{i}$. Therefore, we can replace 
$\frac{\partial \bm{n}}{\partial \bm{R}_{i}^{a}}$ by $-\frac{\partial \bm{n}}{\partial \bm{r}^{a}}$, so
we simply get the topological charge associated to a Skyrmion. With this approximation, the reduced
symplectic form reads:
\begin{equation}
    \Omega_{\mathrm{red}}(\{\delta \bm{R}_{i}^{(1)}\},\{\delta \bm{R}_{i}^{(2)}\}) =\frac{\tau}{l_B^2}\sum_{i=1}^{N_S}
    (\delta \bm{X}_{i}^{(1)} \, \delta \bm{Y}_{i}^{(2)} - \delta \bm{X}_{i}^{(2)} \, \delta \bm{Y}_{i}^{(1)})
\end{equation}
Hamiltonian equations for motion for the skyrmions take then the familiar form
 \begin{equation}
     \dot{X}_i = \tau\frac{l_B^2}{\hbar} \frac{\partial H}{ \partial Y_i},\,\,\,\,\,
     \dot{Y}_i = -\tau \frac{l_B^2}{\hbar} \frac{\partial H}{ \partial X_i}\\
     \label{Hamilton_eq_Skyrmions}
\end{equation}
As for the magnon, it will be convenient to use $l_B$ as the unit length for Skyrmion positions, so that
$l_B$ disappears in these evolution equations.
We emphasize the qualitative difference between the magnon and the Skyrmions from the viewpoint of their
symplectic forms. Because the magnon is a charge neutral object, its phase-space is four-dimensional,
described by the $\bm{r},\bm{p}$ coordinates. By contrast, Skyrmions carry one unit charge, so the corresponding phase-space is two-dimensional and is identified to the physical plane for $\bm{R}_i$.

To summarize, we have reduced the infinite-dimensional phase-space of the non-linear sigma model to a
phase-space of dimension $4+2N_S$ for one magnon coupled to $N_S$ Skyrmions. The Hamiltonian generating the
time-evolution of this system is $H=H_{\mathrm{mag}}+V_{\mathrm{el}}$, with $H_{\mathrm{mag}}$ given by Eq.~(\ref{eq:H_magnon})
and $V_{\mathrm{el}}$ by Eq.~(\ref{eq:V_elastic}).
The corresponding equations of motion for
the magnon are given by~(\ref{Hamilton_eq_magnons}) and those for Skyrmions by~(\ref{Hamilton_eq_Skyrmions}).
We note that when Skyrmions are moving, in particular as the result of their interaction with an
incoming magnon, there is an induced electromotive force acting on the magnon, since it experiences a time-dependent
effective vector potential. In the numerical simulations inspired by the present model, we have neglected these
induction phenomena, on the basis that Skyrmions are slowly moving. It is an interesting direction for future work to
explore the consequences of (relaxing) this simplifying assumption.

\section{Effects of electric field gradients in the dipole picture}

Let us return to the situation considered in Fig.~E.D.1 of a fully polarized
Landau level in the presence of spatially varying electric fields. A magnon dipole trajectory is described by the effective Hamiltonian~(\ref{eq:H_magnon}) without the effective vector potential $\bm{A}$.
Motivated by the geometry of the experiment reported in Fig.~E.D.1, we assume that
there exists an in-plane electric field $E_x$ across the junction. To simplify the analysis, 
we replace the junction by an infinite vertical strip into which the electrostatic potential
$V_{\mathrm{ext}}$ is confined and where it depends only on the $x$ coordinate. The resulting
translation symmetry along the $y$ direction ensures that $p_y$ is conserved. The projection
of the magnon motion along the $x$ axis is given by the effective Hamiltonian, where 
$p_y$ enters as a parameter:
\begin{equation}
H_{p_y}(x,p_x) = \frac{\omega_{\mathrm{int}}}{2} p_{x}^{2} - p_y\,\partial_x V_{\mathrm{ext}}(x)  
\label{eq:H_magnon_x_projected}
\end{equation}
Let us assume that $\partial_x V_{\mathrm{ext}}$ is positive in the interval
$x_L < x < x_R$ and that it vanishes outside this interval. The dynamics depends
crucially on the sign of $p_y$. If $p_y$ is positive, the $x$ coordinate sees
an effective attractive potential well in the intermediate region. If a magnon is launched
from the left with a positive velocity $\dot{x}=\omega_{\mathrm{int}}\,p_x$,
it crosses the central region where it is accelerated and it
recovers the initial value of $p_x$ for $x>x_R$. By contrast, if $p_y$ is negative,
the $x$ coordinate is subjected to a localized repulsive potential. Two different behaviors
can occur. If $|p_y|<p_{y*}(p_x)$ with $p_{y*}(p_x)=\omega_{\mathrm{int}}\,p_{x}^{2}/2 \mathrm{Max}(\partial_x V_{\mathrm{ext}})$,
the magnon still crosses the central interval, where it slows down. But when
$|p_y|>p_{y*}(p_x)$, the magnon bounces back. 

To complete the description of classical magnon trajectories, the motion along the
$y$ direction is governed by the equation:
\begin{equation}
\dot{y} = \omega_{\mathrm{int}}\,p_y - \partial_x V_{\mathrm{ext}}(x)    
\end{equation}
Let us consider the total shift $\Delta y$ in the $y$ coordinate between two trajectories
with the same initial conditions, one in the presence and the other in the absence of the
external potential. With our choice of positive $\partial_x V_{\mathrm{ext}}$, we see
that $\Delta y$ is always negative. Clearly, it accumulates only when the $x$
coordinate satisfies $x_L < x < x_R$. At large $|p_y|$, $\Delta y$ goes to zero.
In the case of large and positive $p_y$, this is due to the fact that the magnon undergoes a strong acceleration in the central region, in which it spends therefore a very small time.
For large and negative $p_y$, the magnon bounces almost immediately after reaching the
left side of the central region. For intermediate values, we see that it diverges
at the limiting value $p_{y*}(p_x)$ separating the two regimes characterized either by transmission or by bounce. An illustration of these trajectories is shown on Figure~E.D. 4.

\section{Magnon reflection and adsorption at boundaries}
The above model can be slightly modified to understand the effect of Hall droplet boundaries
on magnon trajectories as described by the dipole model. Consider an infinite vertical boundary
located at $x=0$. We model this situation by assuming that $V_{\mathrm{ext}}(x)=0$ for
$x \leq 0$ and $V_{\mathrm{ext}}(x)$ increases monotonously for $x \geq 0$. Assuming a smooth
confining potential we also assume that exactly the same features hold for the first derivative
$\partial_x V_{\mathrm{ext}}(x)$, at least up to some positive value $x_R$. 
This interface is well-known to host a chiral edge mode characterized by a local velocity
$\dot{y}=-\partial_x V_{\mathrm{ext}}(x)$ in units of the magnetic length. Note that if a dipole is
created near the interface, the drift velocities of the particle and of the hole constituents would 
be given by this value if we could ignore the Coulomb interaction between them. This interaction
is captured by the $\frac{\omega_{\mathrm{int}}}{2}\,{\mathbf{p}^2}$
term in the dipole effective Hamiltonian. Let us assume that a magnon is launched from the left
towards the interface, so that $p_x$ is positive. As we have just seen, the previous model
predicts that the magnon bounces with nearly specular reflection (in the limit of a fast rising
potential) only when $p_y$ is negative. The positive $p_y$ case seems at first glance surprising,
because the model predicts that the magnon could travel arbitrary far to the right, therefore escaping 
the confining potential! Of course, the effective Hamiltonian~(\ref{eq:H_magnon}) holds only inside
regions where a hole can be created, so that the local filling factor in the absence of magnons
should remain close to unity. At the edge of the droplet, the electronic density quickly falls to zero,
so the magnon has to stick to the edge, when it is not bouncing. After this adsorption event, 
both the particle and the hole component drift along the edge with the same velocity. Their relative
distance is fixed by the internal Coulomb energy of the dipole. 

To summarize this discussion, we can recast it in the following physical terms. 
To predict whether a magnon impinging on a boundary will be reflected or adsorbed, we have to 
examine the relative orientation of the tangential velocity of the magnon with the 
local particle flow along the edge induced by the confining potential. If these two orientations
coincide (as for $p_y <0$ in the previous paragraph), the particle piece hits the boundary before
the hole, so that the dipole orientation changes suddenly and the dipole is reflected. 
If these two orientations are opposite, the dipole aligns with the boundary without changing
the orientation of its projection along the edge, so it is adsorbed.

\section{Phenomenological model involving charged particle in a flux-tube array }

In the previous section, we have presented a complete semiclassical description of the magnon-skyrmion crystal junction problem, deriving equations of motion for both the magnons and the skyrmions. We expect such an analytical treatment to be a starting point for several other such similar situations of magnons interacting with quantum Hall type insulators with non-trivial topology.

In the absence of exact information about junction parameters, we have also postulated several forms of the electrostatic potentials and effective magnetic fields based on physical arguments and approximations. Here we shall summarize the full phenomenological setup and explain the numerical procedure for obtaining the magnon trajectories and variance.

Based on the above sections illustrating the equivalence of different microscopic pictures and appropriate semiclassical limits, here we present the details of the simplest phenomenological model which qualitatively captures the sharp noise feature across all detectors in narrow voltage windows. We show that shaking of the skyrmion crystal by the magnons, when the crystal softens, is the source for the linearly increasing (with voltage) variance in the non-local magnon signal across all detectors.

The crucial ingredients of the phenomenological model are:

i) the two confining potentials along $x$ and $y$ which constitute $V_{ext} (\bm{r})$. For the former we use a quadratic potential, $V_{cx} = K_x x^2/(1+(x/x_0)^2)$ for the junction width, which smoothly saturates to a constant, $K_x x_0^2$, away from the junction for $x \gg x_0$. For the latter, we use a flat potential $V_{cy} = 0$, along the junction region and a hard-wall, $V_{cy} = K_y (|y|-y_J)^4$ form, beyond the junction.

ii) The electrostatic potential induced in the junction region by the skyrmion crystal, for which we use the form presented in the second term of eq. (\ref{vlocphenom}).

iii) The effective magnetic field induced by the skyrmion crystal's modulated topological charge density. We take this to have the $B(\bm{r}-\bm{R})$ to have the form presented in eq. \ref{vecphenom}.

iv) The repulsive nearest neighbour inter-skyrmion potential, which we take to be of the form in the second term of  eq. \ref{eq:V_elastic}. Note that the first term is the same $V_{ext}$ as in point (i) above, but now for the skyrmion coordinates.

These ingredients are all the inputs required to numerically solve the equations of motion for the magnon (eq. \ref{mageqm}) and the skyrmions (eq. \ref{Hamilton_eq_Skyrmions}). Let us write these down more explicitly to clearly see the different terms induced by the different factors. First, we focus on the skyrmions, the Hamiltonian for which is given by $H = H_{mag} + V_{el}$. While the second term will give us the standard potential derivatives, the dependence of the first term on the skyrmion coordinates results in the magnon recoil induced collective dynamics. Recall that skyrmion centre coordinates are denoted as $X_i,Y_i$ and magnon position and momenta are denoted as $x,y$ and $p_x,p_y$ respectively. We can write
\begin{multline}
        \dfrac{\partial H_{mag}}{\partial Y_i} = \frac{\omega_{int}}{2} \bigg[ -2(p_x-A_x) \frac{\partial A_x}{\partial Y_i}-2(p_y-A_y) \frac{\partial A_y}{\partial Y_i}\bigg] + (p_x-A_x)\partial_{Y_i} \partial_y V_{loc} - (\partial_y V_{loc})(\frac{\partial A_x}{\partial Y_i}) \\ -(p_y-A_y)\partial_{Y_i} \partial_x V_{loc} + (\partial_x V_{loc})(\frac{\partial A_y}{\partial Y_i}) \\
        \dfrac{\partial H_{mag}}{\partial X_i} = \frac{\omega_{int}}{2} \bigg[ -2(p_x-A_x) \frac{\partial A_x}{\partial X_i}-2(p_y-A_y) \frac{\partial A_y}{\partial X_i}\bigg] + (p_x-A_x)\partial_{X_i} \partial_y V_{loc} - (\partial_y V_{loc})(\frac{\partial A_x}{\partial X_i}) \\ - (p_y-A_y)\partial_{X_i} \partial_x V_{loc} + (\partial_x V_{loc})(\frac{\partial A_y}{\partial{X_i}})  
\end{multline}

This set of formulae, combined with the standard potential derivatives of $V_{el}$ form the equations of motion for the skyrmions. To proceed with the calculations, we choose the Landau gauge with $A_x = 0$ and $A_y = \sum_{i=1}^{N_S} (x-X_i) B_{\bm{r'R_i}}  $ where $B_{\bm{r'R_i}} = 4a^2/(|\bm{r'}-\bm{R_i}|^2+a^2)^2$ as in eq. (\ref{vecphenom}). As a consequence we get 
\begin{equation}
\begin{aligned}
    \dfrac{\partial A_y}{\partial x} &= \sum_{i=1}^{N_S} B_{\bm{r}'\bm{R}_i} = B ;  \dfrac{\partial A_x}{\partial y} =\dfrac{\partial A_x}{\partial X_i } =   \dfrac{\partial A_x}{\partial Y_i} =0 \\
      \dfrac{\partial A_y (x,y,\bm{X},\bm{Y})}{\partial X_i} &=  - B_{\bm{r},\bm{R}_i}  \\ 
      \dfrac{\partial A_y (x,y,\bm{X},\bm{Y})}{\partial Y_i} &= (x-X_i)\dfrac{\partial B_{\bm{r},\bm{R}_i}}{\partial Y_i} = \dfrac{16a^2 (x-X_i)(y-Y_i)}{(|\bm{r}-\bm{R}_i|^2+a^2)^3}
\end{aligned}
\end{equation}
In a similar vein we rewrite the magnon equations of motion.
\begin{equation}
    \begin{aligned}
        \ddot{x} &= (\omega_{int}B + \partial^2_x V_{loc}+\partial^2_y V_{loc})\dot{y} + (\partial_y V_{loc})(\partial_x \partial_y V_{loc}) + (\partial_x V_{loc})(\partial^2_x V_{loc}) \\
        \ddot{y} &= -(\omega_{int}B + \partial^2_x V_{loc}+\partial^2_y V_{loc})\dot{x} + (\partial_y V_{loc})( \partial^2_y V_{loc}) + (\partial_x V_{loc})(\partial_y \partial_x V_{loc}) 
    \end{aligned}
\end{equation}
where $B$ is the total magnetic field generated by all the skyrmions. We clearly see that the variation of the electrostatic potential (also modified by the skyrmions) contributes an effective Lorentz force as well as a scalar potential.
The various derivatives of the electrostatic potential felt by the magnons also act as sources of effective magnetic field and scalar potentials as highlighted in the above sections. They are given by 
\begin{equation}
\begin{aligned}
    \partial_x V_{loc} &= \partial_x V_{cx} - \sum_{i=1}^{N_S} \dfrac{(x-X_i)}{(|\bm{r}-\bm{R}_i|^2+b^2)^{3/2}}; \; \partial_y V_{loc} = \partial_y V_{cy} - \sum_{i=1}^{N_S} \dfrac{(y-Y_i)}{(|\bm{r}-\bm{R}_i|^2+b^2)^{3/2}} \\
    \partial_{X_i}\partial_x V_{loc} &= \dfrac{1}{(|\bm{r}-\bm{R}_i|^2+b^2)^{3/2}} - \dfrac{3(x-X_i)^2}{(|\bm{r}-\bm{R}_i|^2+b^2)^{5/2}}; \; \partial_{Y_i}\partial_y V_{loc} = \dfrac{1}{(|\bm{r}-\bm{R}_i|^2+b^2)^{3/2}} - \dfrac{3(y-Y_i)^2}{(|\bm{r}-\bm{R}_i|^2+b^2)^{5/2}}; \\
  \partial_{Y_i}\partial_x V_{loc}  &=  \partial_{X_i}\partial_y V_{loc} = \dfrac{-3(x-X_i)(y-Y_i)}{(|\bm{r}-\bm{R}_i|^2+b^2)^{5/2}}
\end{aligned}
\end{equation}

We find it quite striking that the magnon-skyrmion crystal model, when captured within this non-linear sigma model so clearly elucidates the interplay of topology, geometry and collective dynamics. We hope that such analytical analysis serves as a starting point for similar magnon based probes of other topological phases of matter.

We numerically solve the above equations of motion to obtain the magnon and skyrmion trajectories. We consider the magnons to be scattered, one by one, with an initial velocity $\bm{v}$ and incidence angle $\theta_i$.
These equations form the basis of our phenomenological model, which we then numerically solve using standard symplectic mid-point integration techniques. We start with a five skyrmion, lined along the $y$-axis range in the central region around x = 0.  Now we consider charged particles starting at $x = -L$ and at a random point in the $y = [-4,4]$ range. We do so to simulate emission from the central detector at the end of the left slab, as is done for the noise data in the experiment.
For Fig. 4C in the main text, we simulate a total of 10000 magnons starting from the initial conditions described above, with $dt = 0.001$ time step in our midpoint Strang symplectic integration scheme and we vary the number of steps per pass to simulate the varying magnon flux. To obtain the variance data shown in Fig. 4C, we define a suitable unit of time and simply obtain the number of charged particles that hit each of the detectors (three different y coordinate ranges at x = +L) in that unit of time. This is the magnon count, also illustrated in Fig 3D and E. Further,  we model the reduced stiffness by reducing the coefficient of the x-axis confining potential $K_x$. As shown in Fig.~E.D. 5, a reduction in stiffness drastically increases the noise in the skyrmion positions.

Our results lend qualitative support to the picture that, dynamics of the skyrmion crystal induced by the impinging magnons is the cause for the sharp noise observed in the experiment at periodic voltages corresponding to the discrete process of adding an extra skyrmion. Within our simple model this can be illustrated clearly, as in Fig.~E.D. 5, by explicitly looking at the trajectories of the flux tubes and their deflection when a charged particle passes by.  

\clearpage

%% file: SI.tex
\title{Supplementary Information for: Magnons reveal topology and dynamics of a skyrmion crystal}

\author{Rapha\"el Ayache}
\thanks{These authors contributed equally}
\affiliation{SPEC, CEA, CNRS, Université Paris-Saclay, CEA Saclay, 91191 Gif sur Yvette Cedex
France}

\author{Nilotpal Chakraborti}
\thanks{These authors contributed equally}
\affiliation{TCM Group, Cavendish Laboratory, University of Cambridge, Cambridge CB3 0HE, United Kingdom}

\author{Manabendra Kuiri}
\thanks{These authors contributed equally}
\affiliation{SPEC, CEA, CNRS, Université Paris-Saclay, CEA Saclay, 91191 Gif sur Yvette Cedex
France}
\affiliation{Department of Physics, Birla Institute of Technology and Science, Pilani, Hyderabad Campus, Jawahar Nagar, Kapra Mandal, Medchal District, Telangana 500078, India}

\author{Quentin Benichou}
\affiliation{SPEC, CEA, CNRS, Université Paris-Saclay, CEA Saclay, 91191 Gif sur Yvette Cedex
France}

\author{Antonio Lacerda-Santos}
\affiliation{SPEC, CEA, CNRS, Université Paris-Saclay, CEA Saclay, 91191 Gif sur Yvette Cedex
France}

\author{Lilian Seyve}
\affiliation{SPEC, CEA, CNRS, Université Paris-Saclay, CEA Saclay, 91191 Gif sur Yvette Cedex
France}

\author{Himadri Chakraborti}
\affiliation{SPEC, CEA, CNRS, Université Paris-Saclay, CEA Saclay, 91191 Gif sur Yvette Cedex
France}
\affiliation{Laboratory of Atomic and Solid State Physics, Cornell University, Ithaca 14850, NY, USA}

\author{L\'eo Pugliese}
\affiliation{SPEC, CEA, CNRS, Université Paris-Saclay, CEA Saclay, 91191 Gif sur Yvette Cedex
France}

\author{Kenji Watanabe}
\affiliation{Research Center for Functional Materials, National Institute for Materials Science, Japan}

\author{Takashi Taniguchi}
\affiliation{International Center for Materials Nanoarchitectonics, National Institute for Materials Science, Japan}

\author{Roderich Moessner}
\affiliation{Max-Planck Institut für Physik komplexer Systeme, Nöthnitzer Straße 38, Dresden 01187, Germany}

\author{Cosimo Gorini}
\affiliation{SPEC, CEA, CNRS, Université Paris-Saclay, CEA Saclay, 91191 Gif sur Yvette Cedex
France}

\author{Beno\^it Doucot}
\affiliation{LPTHE, UMR 7589, CNRS and Sorbonne Université, 75252 Paris Cedex 05, France}

\author{Preden Roulleau}
\email{preden.roulleau@cea.fr}
\affiliation{SPEC, CEA, CNRS, Université Paris-Saclay, CEA Saclay, 91191 Gif sur Yvette Cedex
France}

\maketitle

\renewcommand{\thefigure}{S.\arabic{figure}}

\section{Device Geometry}
\begin{figure}[h]
    \centering
    \includegraphics[width=0.85\linewidth]{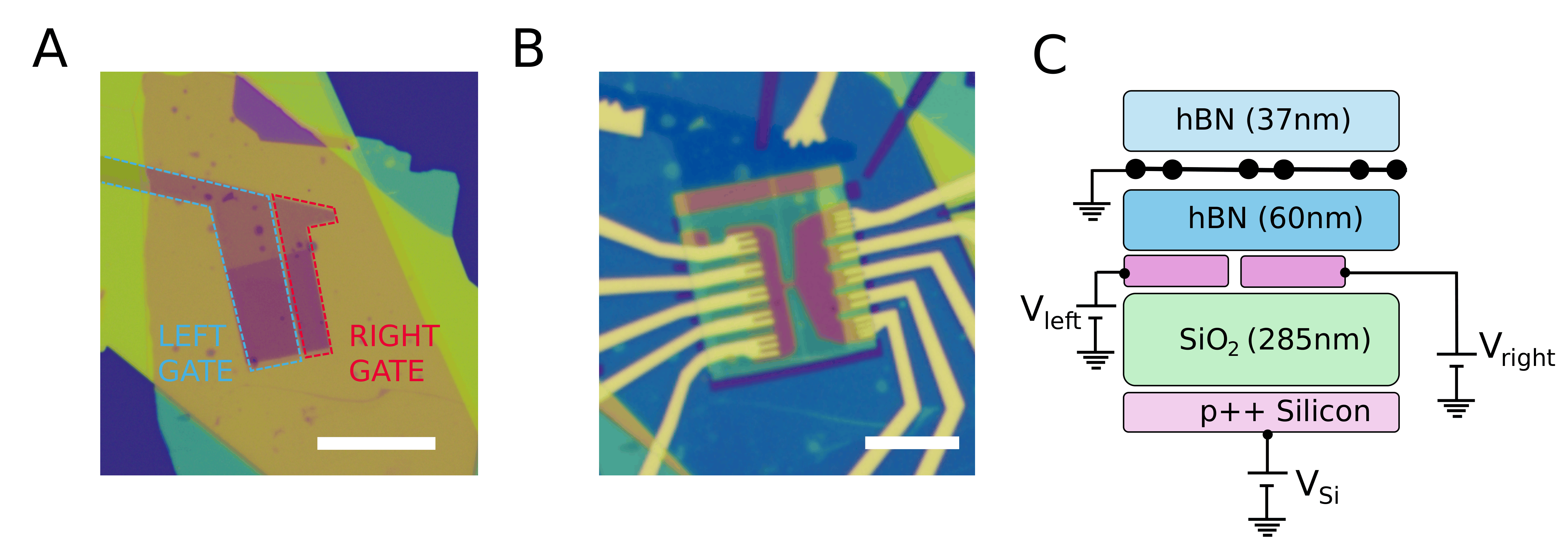}
    \caption{
        (a) Optical image of the hBN/graphene/hBN/graphite stack (top to bottom). 
        The bottom graphite was pre-patterned into two gates shown in dashed red (left gate) and blue (right gate), respectively. Scale bar is 5~$\mathrm{\mu m}$. 
        The separation between left and right gates is $\sim$70~nm. 
        (b) Optical image of the final device. Scale bar is 5~$\mathrm{\mu m}$. 
        (c) Cross-sectional schematic of the device. 
        right-gate voltage ($V_{\mathrm{right}}$) and left-gate voltage ($V_{\mathrm{left}}$) 
        were applied to the two graphite gates as shown. 
        These two gates are separated by $\sim$70~nm.
    }
    \label{fig:S1}
\end{figure}

\begin{figure}[h]
    \centering
    \includegraphics[width=0.75\linewidth]{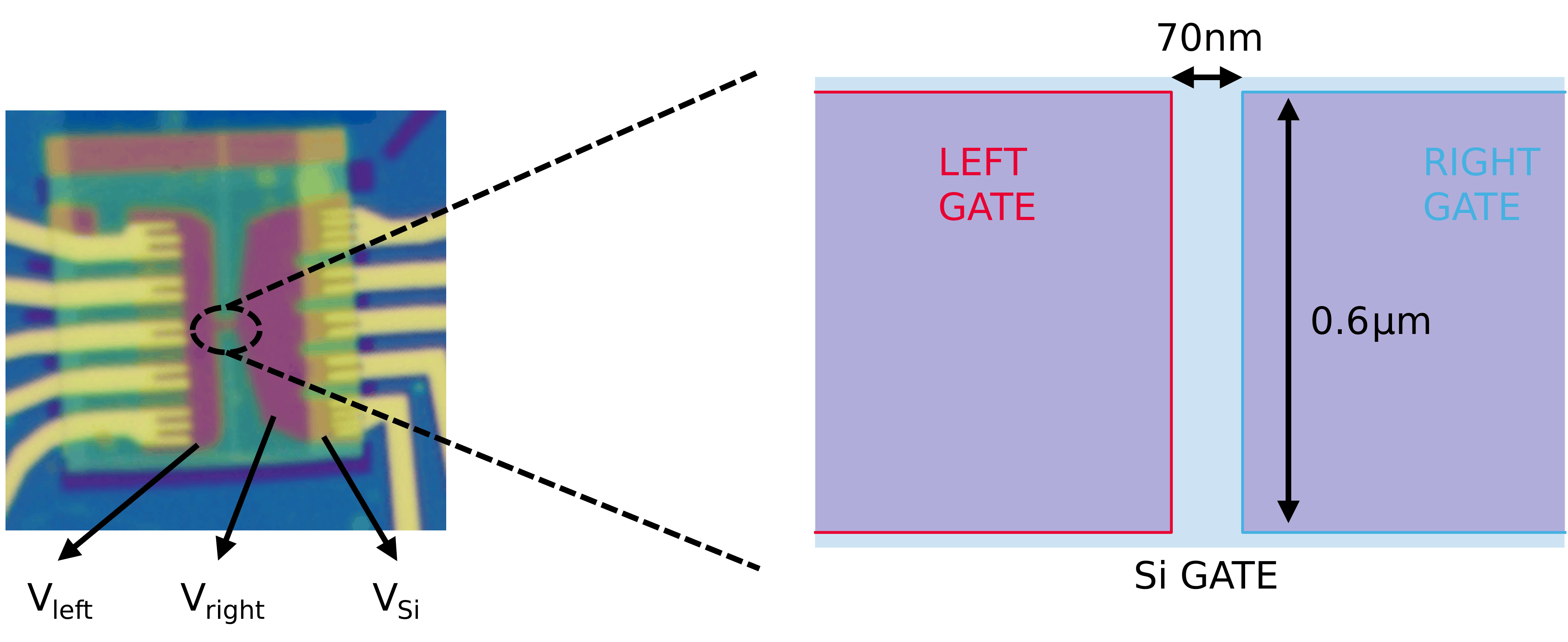}
    \caption{
        Optical image of the device showing the details of the constriction 
        and the gate configuration. The width of the graphene channel and the gates 
        is 0.6~$\mu$m, and the distance between left and right gates is 70~nm.
    }
    \label{fig:S2}
\end{figure}

Supplementary Fig.~\ref{fig:S1} and Fig.~\ref{fig:S2} show the van der Waals heterostructure used to create our device before and after fabrication, highlighting the details of the patterned bottom graphite gate. 

\section{Electrical Transport Measurements}

Transport measurements were conducted using an Oxford Instruments Proteox dilution refrigerator with a base temperature of approximately 10~mK and a magnetic field of up to 14~T applied perpendicular to the sample plane. Standard low-frequency lock-in techniques were employed to probe the electrical response of the device. An excitation current in the range of 1--5~nA was applied using a Stanford Research Systems SR830 lock-in amplifier, with modulation frequencies varied between 7~Hz and 217~Hz to optimise signal to noise conditions across different measurement regimes. The output voltage signal was amplified using a low-noise, high-impedance preamplifier (input impedance $\sim$100~M$\Omega$). All electrical lines were filtered using low-pass RC filters, and were thermally anchored at multiple stages within the cryostat to suppress spurious heating and minimise electronic noise.
 
\section{Quantum Hall Transport Characteristics}
 
\begin{figure*}[h]
    \centering
    \includegraphics[width=\textwidth]{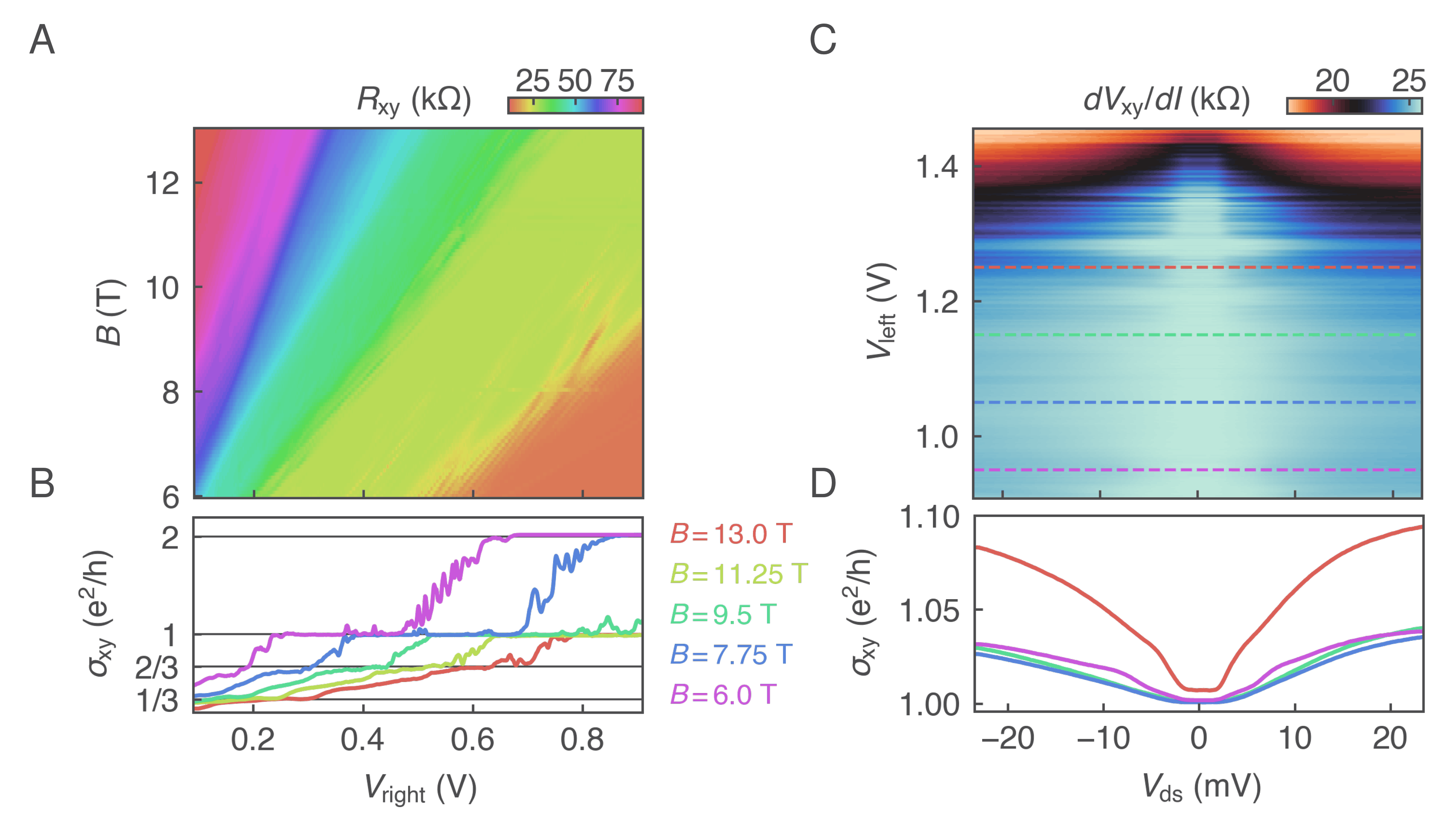}
    \caption{
        \textbf{(a)} Landau fan of the two-terminal Hall resistance $R_{xy}$ measured from 6 to 13~T while sweeping the right gate voltage. A contact resistance of 750~$\mathrm{\Omega}$, corresponding to the series resistance of the device contacts, has been subtracted. Well-quantized plateaux are observed at $\sigma = e^{2}/h$, $2e^{2}/h$, $1/3, e^{2}/h$, and $2/3, e^{2}/h$, demonstrating the high sample quality and the presence of robust integer and fractional quantum Hall states.
        \textbf{(b)} Line cuts of $R_{xy}$ at fixed magnetic fields (6, 7.75, 9.5, 11.25, and 13~T), illustrating the persistence and sharpness of both integer and fractional plateaux across the full field range.
        \textbf{(c)} Breakdown of the $\nu = 1$ quantum Hall state as a function of source–drain bias. A small reduction of 2.5–8\% at $V_{\mathrm{ds}} = 20$~mV signals the onset of dissipative processes, preceding the full collapse of the plateau at higher bias.
        \textbf{(d)} Representative bias-dependent cuts illustrating the full breakdown of the $\nu = 1$ plateau.
    }
    \label{fig:QH_characterization}
\end{figure*}

To assess the quality of the device and establish the robustness of the quantum Hall regime in our junction geometry, we first characterize the Hall response as a function of magnetic field and carrier density. Figure~\ref{fig:QH_characterization}a shows a Landau fan of the transverse resistance $R_{xy}$, measured in a two-terminal configuration while sweeping the right gate voltage between 6 and 13~T. A constant series resistance of 750~$\mathrm{\Omega}$—corresponding to the contact resistance of the measurement line—has been subtracted from all traces. Well-developed quantum Hall plateaux appear at $\sigma_{xy} = e^{2}/3h$, $2 e^{2}/3h$, $e^{2}/h$, and $2e^{2}/h$. The observation of fractional plateaux demonstrates the high mobility and low disorder of the sample. Figure~\ref{fig:QH_characterization}b displays representative line cuts of $R_{xy}$ at fixed magnetic fields (6, 7.75, 9.5, 11.25, and 13~T).

To further probe the stability of the integer states, we examine the breakdown of the $\nu = 1$ quantum Hall plateau. Figure~\ref{fig:QH_characterization}c shows the bias dependence of the $\nu = 1$ state. The plateau remains quantized up to moderate bias, but a small decrease of 2.5–8\% is observed at $V_{\mathrm{ds}} = 20$~mV, signaling the onset of dissipative processes prior to full breakdown. Representative cuts in Fig.~\ref{fig:QH_characterization}d illustrate the breakdown of the $\nu = 1$ state. The well-quantized Hall plateaux indicate that the bulk is largely incompressible over the measured gate and field range, providing a clean environment for magnon propagation.

Together, these results confirm that the device enters a well-developed integer quantum Hall regime with robust plateaux, demonstrating a largely incompressible bulk that provides a clean platform for investigating magnon propagation and electrically tunable scattering across the junction.

\section{Magnon emission and detection mechanism}
 
\begin{figure}[ht]
    \centering
    \includegraphics[width=0.85\linewidth]{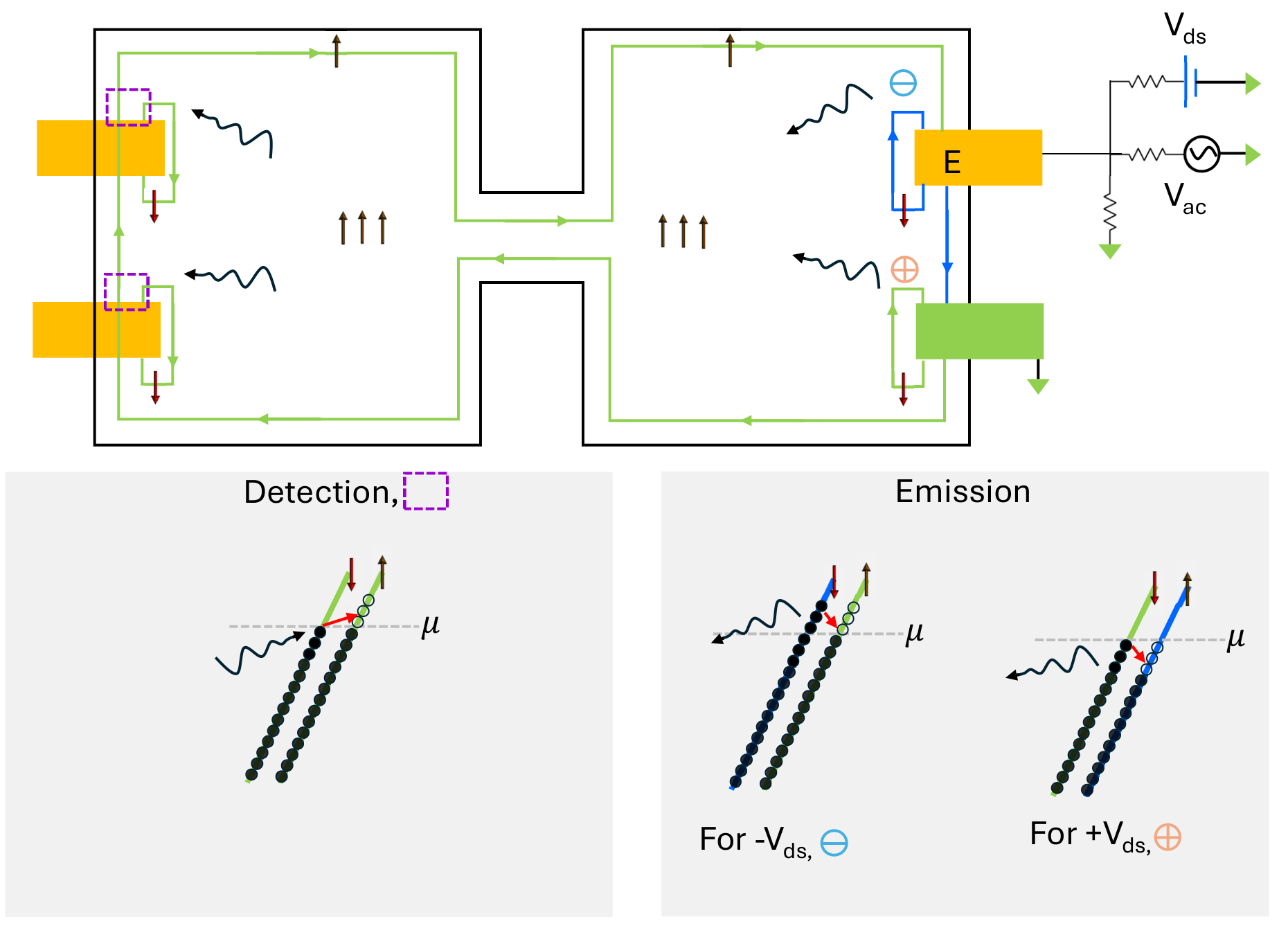}
    \caption{
        \textbf{Top panel}: Schematic of the measurement setup used in this experiment. \textbf{Bottom panel}: Edge-state illustration of magnon emission and absorption processes. The symbols $\ominus$ and $\oplus$ denote the locations where magnons are emitted for negative and positive $V_{ds}$, respectively. 
    }
    \label{fig:magnon_process}
\end{figure}

A schematic of the electrical pathways for magnon emission and detection is shown in  Figure~\ref{fig:magnon_process}. The sample is first tuned into the $\nu = 1$ quantum Hall ferromagnetic state by applying appropriate voltages to the two halves of the bottom gate at a perpendicular magnetic field of $13$~T. In this regime, a single spin-polarized chiral edge channel runs along the sample boundary. Near the metallic contacts, however, local doping raises the filling factor, creating an additional inner edge channel of opposite spin polarization. Because these two edge modes carry opposite spins, elastic scattering between them is forbidden unless the energy imbalance exceeds the Zeeman energy $E_Z = g \mu_B B$.

A voltage ($V_{ds}$) is applied to a contact denoted as 'Em'. Because only spin-down angular momentum can enter and propagate through the spin-up bulk of the quantum Hall ferromagnet, magnons are generated at the position marked with a minus sign($\circleddash$) when the chemical potential of the inner edge satisfies $\mu \ge E_Z$ with respect to the outer edge. Conversely, for positive $V_{ds}$, magnon creation occurs at the position marked with a plus sign. Once generated, these magnons propagate through the insulating QH ferromagnet and can be absorbed at remote edge channels through the reverse spin-flip process. Such magnon absorption changes the conductance of the distant edge channel, enabling nonlocal detection of magnon transport.

\clearpage
\newpage
\section{Electrostatic simulations}

We have shown that magnons scattering off of a skyrmion crystal formed at the central junction, confined within a smooth potential, explain the experimental data from Figs. 2.D and 2.E.  To validate our model we now investigate
whether the Graphene junction can host such a crystal by performing simulations of the charge density and electrostatic potential profiles.

Electrostatic screening by the electrons in graphene was included at the mean field (Hartree) level by solving the self-consistent quantum-electrostatics (Schrödinger-Poisson) problem within the Thomas-Fermi approximation. The screening-induced electrostatic reconstruction of quantum Hall edge channels has crucial qualitative ramifications and cannot in general be neglected.  The essential point is that the magnetic (cyclotron) energy $\hbar\omega_c$ is much smaller that the typical electrostatic energy scales in a metallic conductor.  That is, the electronic charge distribution at $B=0$, $n_0(\vec{r})$, is hardly modified by a magnetic field, except in the transition regions between Landau levels where screening is weaker.  For details we refer to seminal works identifying the issue \cite{Beenakker1990, Chang1990} and focusing on electrostatic reconstruction in quantum Hall 2DEGs \cite{Chklovskii1992, Chklovskii1993}.  The reconstruction problem was recently revisited \cite{Armagnat2020}, and specifically in graphene Mach-–Zehnder interferometers \cite{Flor2022}. The lifting of the degeneracy of the graphene Landau levels is introduced phenomenologically through the system's density of states.

The device geometry is shown in Fig.~\ref{fig:device}, with the gate separation varying from $50$ nm to $70$ nm. We studied the junction in three different regimes: $(\nu_L, \nu_R) = (1, -2)$, then $(1, 1)$ and $(1 + \epsilon, 1)$ with $\epsilon \in [0, 1]$. In the simulations, we systematically varied the silicon gate voltage from $15$ to $30V$. We find that, for experimentally relevant gate voltages, the junction develops a localized excess-density pocket whose amplitude and spatial extent depend strongly on $\epsilon$. The confinement is maximal near $\epsilon \approx 0.1$ and progressively weakens for larger $\epsilon$.
Skyrmions formed within the pocket are bound to it, and can thus be described as being confined within an effective potential $V_{ext}({\bf r})$ with a smooth $x$-profile.

\subsubsection{The self-consistent Quantum-Electrostatics problem}

First, we formulate the electrostatic problem. Taking the carrier charge as $q=-e$ with $e$ the elementary charge, the electrostatic potential satisfies:

\begin{equation}
   \label{eq:poisson_intro}
   \nabla \cdot \left( \varepsilon(\vec{r}) \nabla U(\vec{r}) \right) =
   e n(\vec{r},\mu) -e n_d(\vec{r});
\end{equation}

with the local chemical potential given by $\mu=\mu(\vec{r})=eU(\vec{r})$, assuming $E_F=0$. Here $\varepsilon(\vec{r})$ is the position-dependent dielectric permittivity, $n(\vec{r}, \mu)$ is the electron density treated quantum mechanically -- in the present case the density will depend on $\vec{r}$ via the local chemical potential $\mu(\vec{r})$ only, see below -- and $n_d(\vec{r})$ denotes any electric charge not treated at the quantum mechanical level.

At zero temperature, the electron density is obtained from the integrated local density of states:

\begin{equation}
   \label{eq:ILDOS}
   n(\vec{r}, \mu) = \int_{-\infty}^{eU(\vec{r})} dE \rho(E)
\end{equation}

with $ \rho(E)$ the density of states of bulk homogeneous graphene

\begin{equation}
   \begin{aligned}
      \label{eq:GTF_DOS}
      \rho(E) = & \frac{2 |E|}{\pi \hbar^2v_f^2}                                                                               & \text{for} \quad B = 0     \\
      \rho(E) = & \frac{1}{2\pi l_B^2} \sum_{n\in \mathbb{Z}}\sum_{p=\pm1, \pm3} \delta\left(E - E_n -\frac{p}{4}E_{ex}\right) & \text{for}  \quad B \neq 0
   \end{aligned}
\end{equation}

Here $E_n = \hbar v_F \text{sgn}(n)\sqrt{2|n|} / l_B$ are the degenerate Landau Levels, with $v_f \approx 10^6 m/s$ the fermi velocity and $l_B = \sqrt{\hbar / eB}$ the magnetic length.  The degeneracy is lifted due to exchange interaction, with a splitting introduced phenomenologically and estimated as $E_{ex}=30meV$.  As shown in Ref.~\cite{Flor2022} this value allows to determine with a $10-20$\% precision the width of Mach--Zehnder interferometers formed at graphene pn-junctions, which is accurate enough for our current purposes.

\begin{figure}
   \includegraphics[width=.6\textwidth]{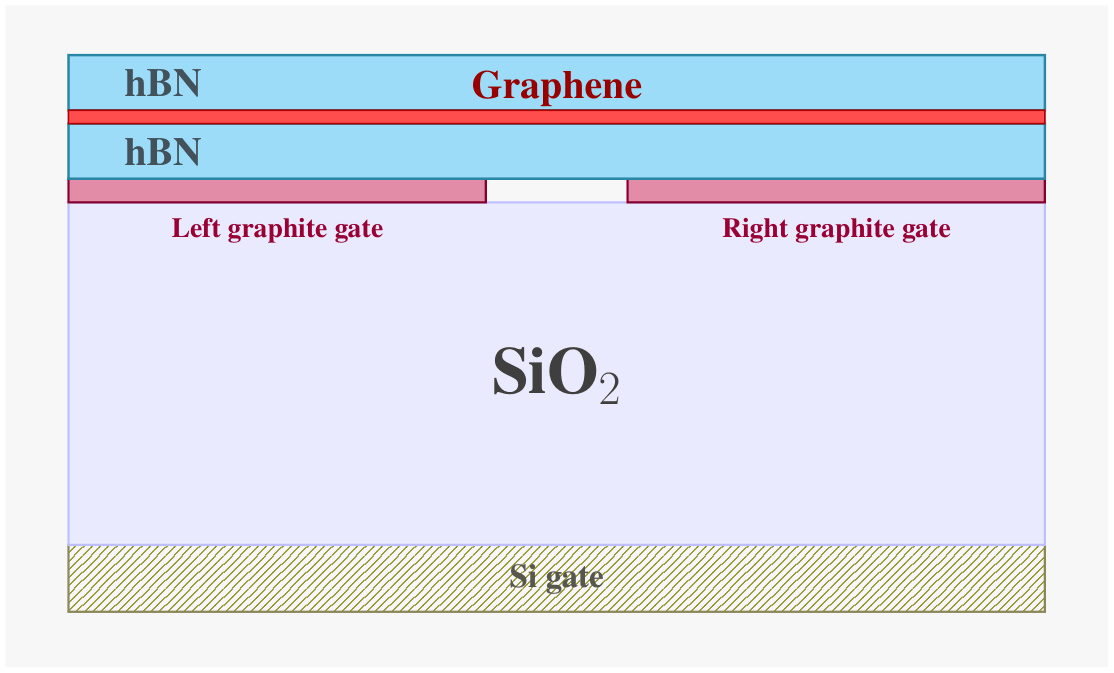}
   \caption{Schematic cross-section of the simulated device geometry.}
   \label{fig:device}
\end{figure}

We consider the device in Fig.~\ref{fig:device} and assume translational invariance along the junction. The simulations are therefore restricted to the 2D plane perpendicular to the junction - as illustrated in Fig.~\ref{fig:device}. At both silicon and graphite gates we apply Dirichlet boundary conditions - thus reducing Eq.\ref{eq:poisson_intro} to the Poisson Equation with $U$ known, $n_d$ unknown and $n=0$. At the dielectric regions ($SiO_2$, hBN and air) we apply Neumann boundary conditions, thus reducing Eq.\ref{eq:poisson_intro} to the Poisson Equation with $n_d = n = 0$ and $U$ unknown. At the graphene region, Eq.\ref{eq:poisson_intro} becomes the non-linear Helmholtz equation :

\begin{equation}
   \label{eq:non_linear_helmholtz}
   \nabla \cdot \left( \varepsilon(\vec{r}) \nabla U(\vec{r}) \right) =
   \int_{-\infty}^{eU(\vec{r})} dE \rho(E);
\end{equation}

where both potential and charge are determined self-consistently. We set $\varepsilon_{hBN} = 3.8 \varepsilon_0$ (see \cite{Laturia2018}), $\varepsilon_{graphene} = 9.32 \varepsilon_0$ and $\varepsilon_{SiO_2} = 3.9 \varepsilon_0$, with $\varepsilon_0$ the vacuum permittivity.

\subsubsection{Technical details}

To solve this problem, we use the open-source software PESCADO. We refer to Refs. \cite{PescadoI, PescadoII} for a description of the algorithms implemented in the package.

Pescado discretizes Eq.~\ref{eq:poisson_intro} using finite volume scheme, which ensures local (and global) charge and flux conservation. The system's boundary condition are applied to the electric flux. At the outer boundaries of the simulation box there is no incoming or outgoing electric flux.

The simulation box is $5000$ nm long along the $\vec{x}$ direction and $1500$ nm tall along the $\vec{z}$ direction, sufficiently large to contain the 2D cross-section shown in Fig.~\ref{fig:device}. We discretize the system using rectangular volume elements with nonuniform sizes.

A fine mesh is used in the region surrounding the junction, with rectangular elements of size $2 nm$ along $\vec{x}$ and $1 nm$ along $\vec{z}$. This fine discretization covers the interval $x\in [-350, 350]nm$, which contains the typical junction width of approximately $130nm$ (extracted from the frequency of Aharonov-Bohm oscillations see section \label{sec:Calib}), and extends vertically from the silicon backgate up to the beginning of the top hBN layer. A medium mesh, with rectangular elements of size $10 nm$ along $\vec{x}$ and $1 nm$ on $\vec{z}$ is used over the full lateral extent $x\in [-2500, 2500]nm$, and extends vertically from the Si gate to the end of the hBN layer. Finally, a coarse mesh with rectangular elements of size $50 nm$ along $\vec{x}$ and $5 nm$ on $\vec{z}$, is used to discretize the air surrounding the device. This coarse mesh allows us to enlarge the simulation domain and capture the relevant electrostatic field lines, while only marginally increasing the mesh points.

To solve the self-consistent problem we use the Piecewise Newton Raphson algorithm implemented by Pescado, see Section 4.2 of Ref.\cite{PescadoII} for details.

\subsubsection{Calibrating the electrostatics: PN junction in the $(\nu_L, \nu_R) = (1, -2)$ regime}
\label{sec:Calib}

\begin{figure}
   \includegraphics[width=\textwidth]{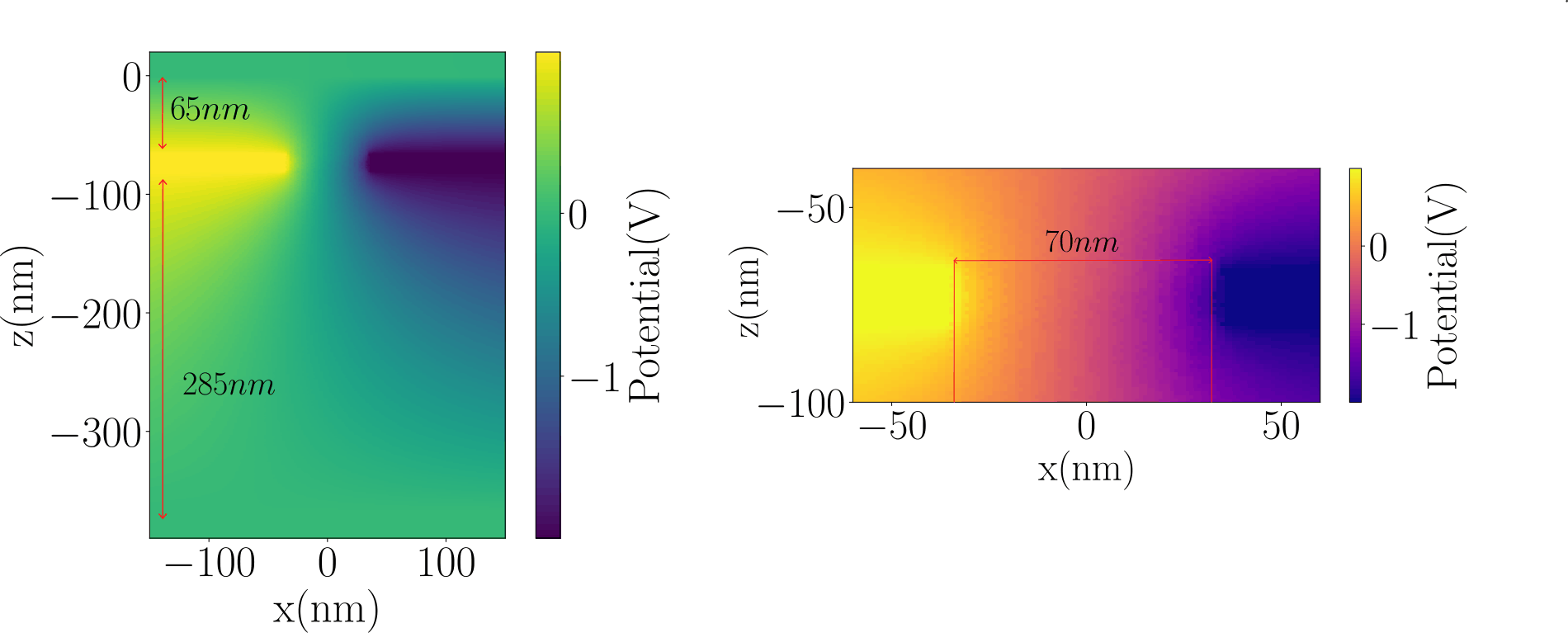}
   \caption{Potential profile for $(\nu_L, \nu_R) = (1, -2)$ with $V_{\text{Si}} = 0V$ and gate separation of $70nm$. On the left we recognize the bottom $hBN$ and $SiO_2$ thickness of respectively $65nm$ and $285nm$. On the right we see the gate separation of $70nm$.}
   \label{fig:Potential_MZI}
\end{figure}

\begin{figure}[h!]
   \includegraphics[width=1\textwidth]{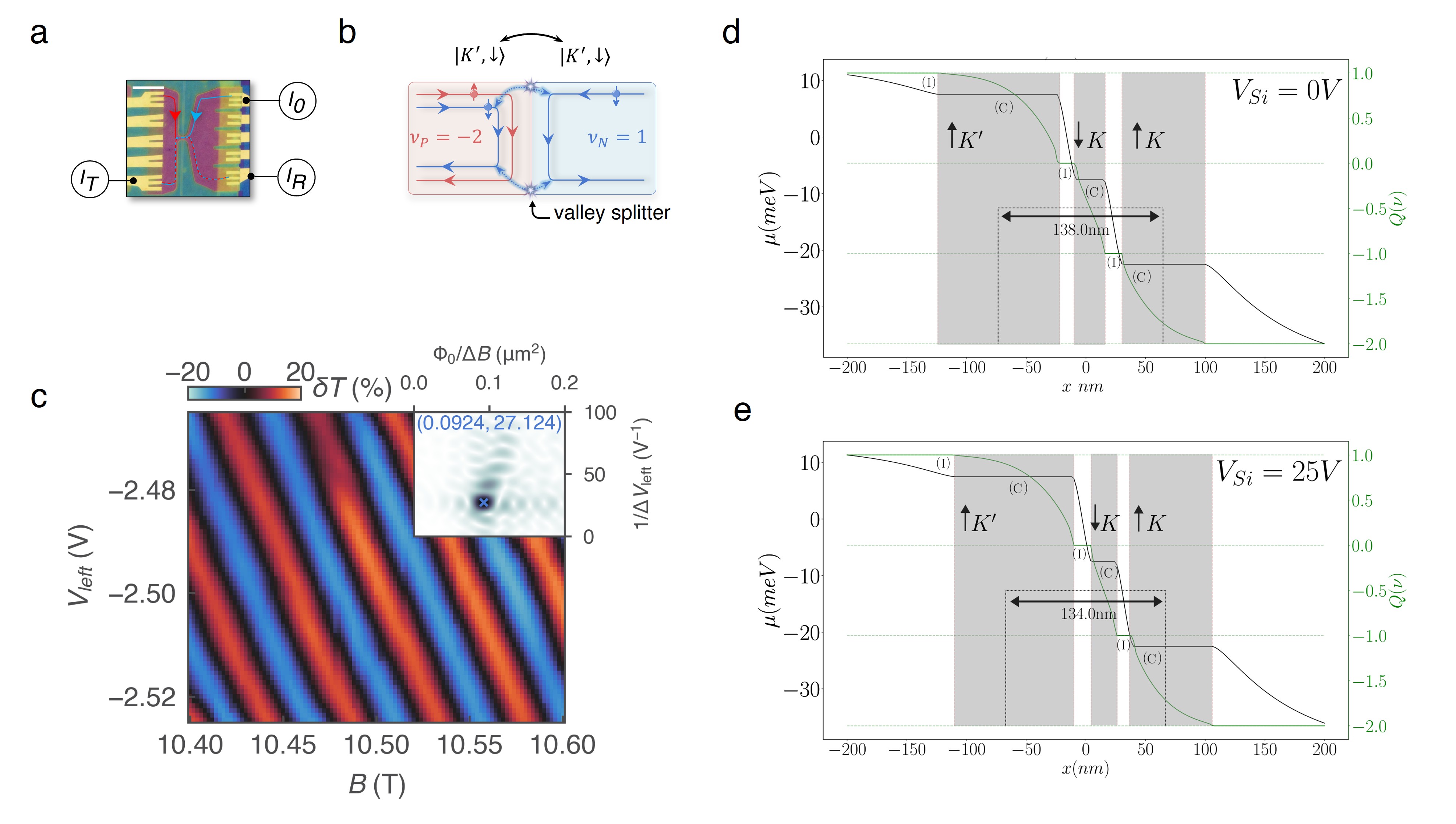}
   \caption{\textbf{a,} Optical micrograph of the device. Red and blue lines denote spin-up and spin-down quantum Hall edge channels, respectively. At the junction, the spin-up channel originating from the left half of the device mixes with the spin-down channel originating from the right half. The resulting transmitted and reflected currents are measured at the corresponding contacts. \textbf{b,} Schematic of the device operated in the $(-2,1)$ filling-factor configuration. In this regime, inter-channel mixing is permitted only along the physical edges, forming two valley splitters that together realize a Mach--Zehnder interferometer. \textbf{c,} Aharonov--Bohm oscillation of the transmitted current as a function of magnetic field and left-gate voltage. The inset shows the corresponding fast Fourier transform. From the oscillation period, an enclosed interferometer area of $\Delta A=\Phi_0/\Delta B = 0.0924~\mathrm{\mu m^2}$ is extracted, corresponding to a separation of $132~\mathrm{nm}$ between the interfering edge channels. \textbf{d-e,} Charge (green) and chemical potential (black) for $(\nu_L, \nu_R) = (1, -2)$ with gate separation of $70$ nm and $V_{\text{Si}} = 0$ V (d) and $V_{\text{Si}} = 20$ V (e). The grey strips are compressible strips and white incompressible regions. The distance of $138$ nm (d) and $134$ nm (e) correspond to the width of the Mach--Zehnder interferometer.}
   \label{fig:SI_MZI_-2_1}
\end{figure}

We start by calibrating the electrostatic simulations against experimental data of the studied device in the configuration of an $PN$-junction type Mach-Zehnder interferometer, see Figure \ref{fig:SI_MZI_-2_1} a-c.

A typical potential profile for the plane of Fig.~\ref{fig:device} is shown in Fig.~\ref{fig:Potential_MZI}. Figure \ref{fig:SI_MZI_-2_1} shows the charge density (green) and potential (black) profiles at the graphene layer for $V_{\text{Si}} = 0~\mathrm{V}$ (top) and $V_{\text{Si}} = 25~\mathrm{V}$ (bottom). The profile can be separated into compressible (grey regions, marked C) and incompressible (white, marked I) regions \cite{Chklovskii1992}. Within the compressible regions the chemical potential is pinned at a Landau level, where the density of states is finite and the charge density can vary, thereby screening the external electrostatic potential -- recall that $\mu(\vec{r})=eU(\vec{r})$ is the total potential including screening.  On the other hand, within the incompressible strips the chemical potential evolves bridging the gap between adjacent Landau levels, where the density of states is zero and the charge density remains fixed. The current-carrying edge channels are associated with compressible regions. Since no spin flip is allowed, only modes with the same spin interfere at the physical edge of the sample. Therefore we define the width of the Mach - Zehnder interferometer as the distance between the centers of the outer compressible strips.

For a gate separation of $70~\mathrm{nm}$, we obtain  $138~\mathrm{nm}$ and $134~\mathrm{nm}$ for $V_{\text{Si}} = 0~\mathrm{V}$ and $V_{\text{Si}} = 25~\mathrm{V}$, respectively. For a gate separation of $50~\mathrm{nm}$, we obtain $129~\mathrm{nm}$ for both values of silicon gate voltage. These values are in good agreement with the experimental width of $132~\mathrm{nm}$, extracted from the period of the Aharonov-Bohm oscillations obtained in the $(\nu_L, \nu_R) = (-2, 1)$ regime for $V_{Si}=32$ V. See Ref.~\cite{Flor2022} for a more detailed comparison between numerically extracted and experimentally determined PN junction widths in similar graphene systems.

\subsubsection{Electrostatics of the junction in the $(1 + \epsilon, 1)$ regime}

\begin{figure}[h!]
   \includegraphics[width=0.5\textwidth]{
      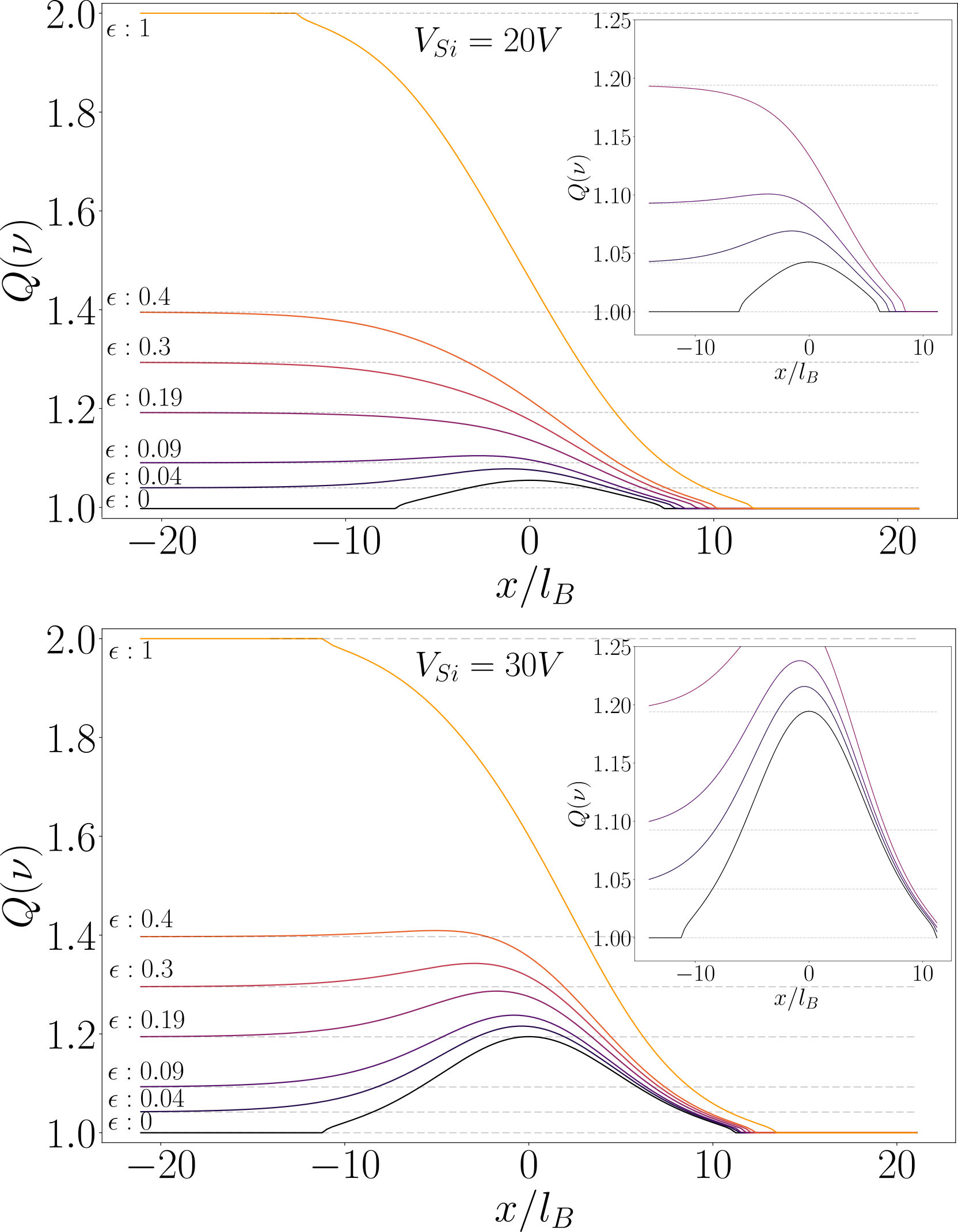}
   \caption{Charge profile as a function of $\nu_L = 1 + \epsilon$ for a gate separation of $70nm$, $V_{\text{Si}} = 20V$ (top) and $V_{\text{Si}} = 30V$ (bottom). The top right insets zoom at the three lowest $ \epsilon $ values. The $x$ axis is normalized by the magnetic length $l_B \approx   \qty{7.1}{\nano\metre}$ at $B=13$T - the size of a skyrmion is on the order of few $l_B$. }
   \label{fig:charge_variation_epsilon}
\end{figure}

We now consider the electrostatics in the regime relevant for the skyrmion - magnon scattering experiments. We normalise all lengths to the magnetic length $l_B \approx   \qty{7.1}{\nano\metre}$ at $B=13$ T.  Figure \ref{fig:charge_variation_epsilon} shows the evolution of the charge density profile as a function of $\nu_L = 1 + \epsilon$ for $V_{\text{Si}} = 20V$ (top) and $V_{\text{Si}} = 30V$ (bottom). A positive silicon gate voltage induces a localized charge-density-peak at the junction, whose amplitude and spatial extent are modulated by $\epsilon$. This charge accumulation arises from electrostatic confinement within the junction, which is a prerequisite to confine charged skyrmions in such a region.  Indeed, the latter  are expected to be favorably confined when their characteristic size -- a few magnetic lengths $l_B$ -- is comparable to the width of the charge-density-peak.  The zoom in Fig \ref{fig:charge_variation_epsilon} focuses on the range $\epsilon < 0.1$, the largest value for which a peak is observed for $V_{\text{Si}} = 20V$. Its spatial extent is clearly visible, and shows that within this parameter range $\epsilon \approx 0.1$ provides the convenient regime for hosting a single skyrmion.

\begin{figure}
   \includegraphics[width=0.5\textwidth]{
      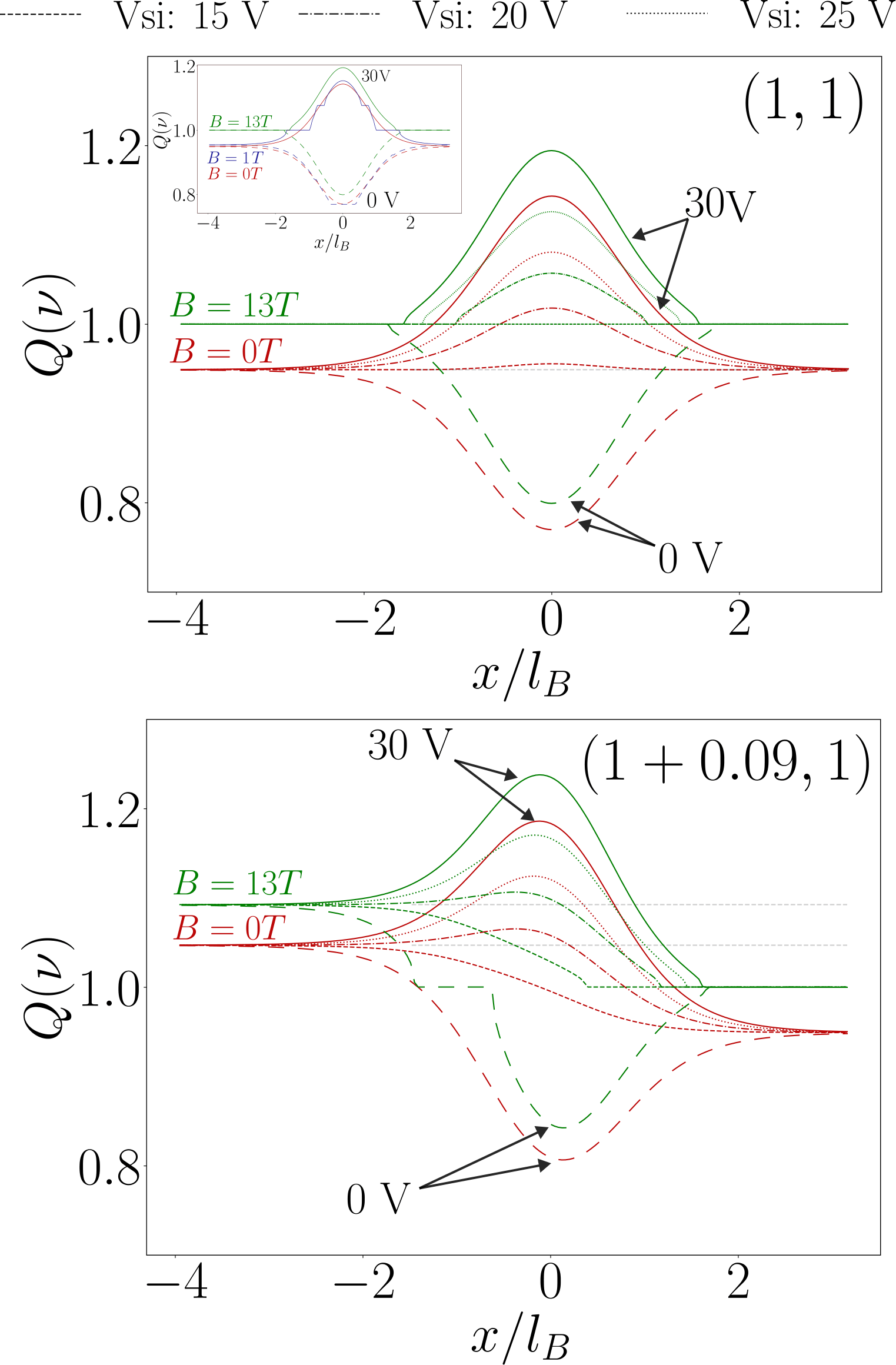}
   \caption{Charge at $B=0T$ (red) and $B=13T$ (green) for $(\nu_L, \nu_R) = (1, 1)$ (top) and $(1 + 0.09, 1)$ (bottom) with gate separation of $70nm$ as a function of the Silicon gate voltage. The zoom on the upper left shows the charge profile for $B=0T$ (red), $B=1T$ (blue) and $B=13T$ (green) for $V_{Si} = 0V$ and $V_{Si} = 30V$. }
   \label{fig:chage_potential_70}
\end{figure}

\begin{figure}
   \includegraphics[width=0.5\textwidth]{
      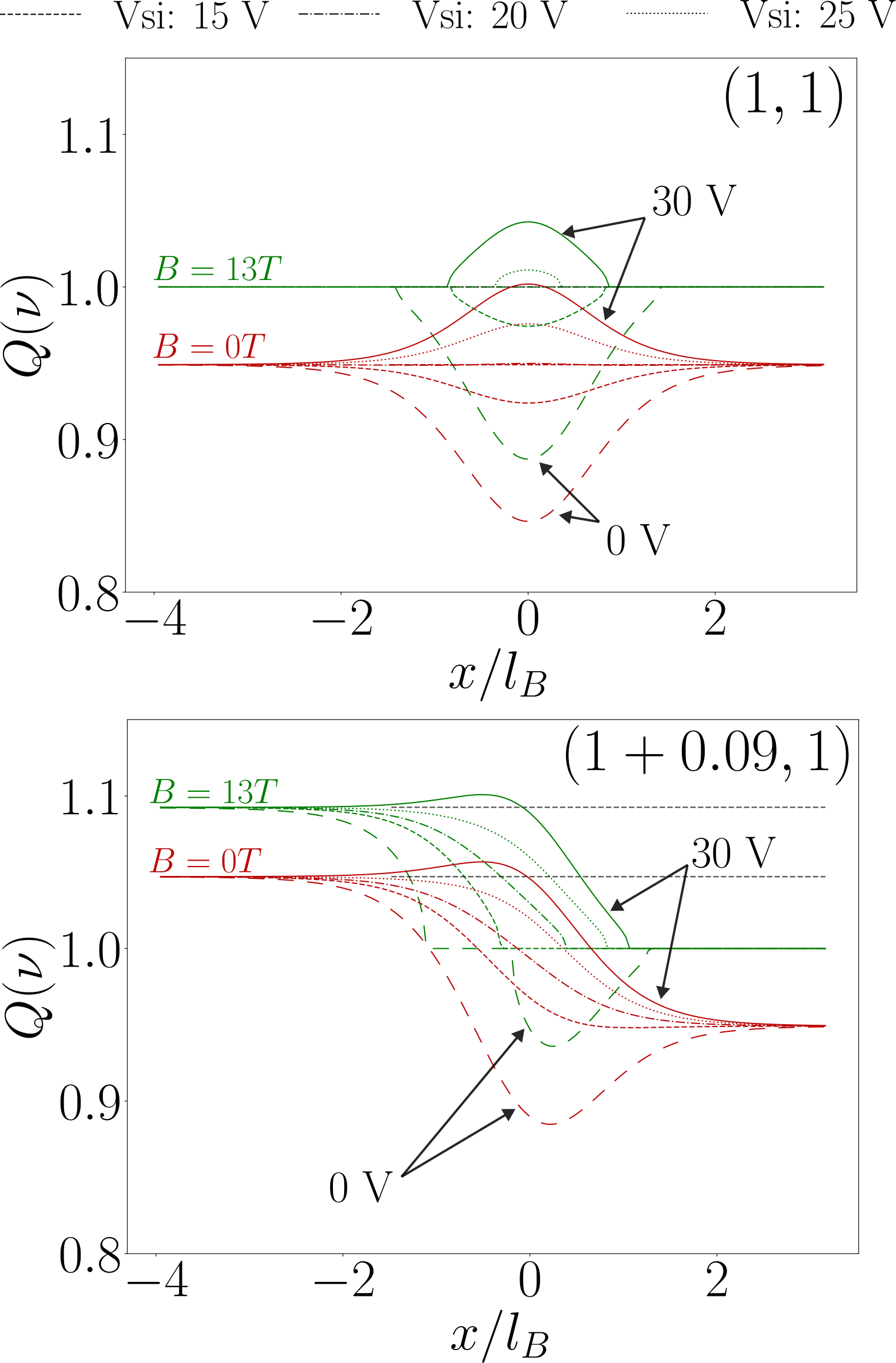}
   \caption{Charge at $B=0T$ (red) and $B=13T$ (green) for $(\nu_L, \nu_R) = (1, 1)$ (top) and $(1 + 0.09, 1)$ (bottom) with gate separation of $50nm$.}
   \label{fig:chage_potential_50}
\end{figure}

Figure \ref{fig:chage_potential_70} shows the charge as a function of magnetic field ($B=0T$ in red, $B=13T$ in green and $B=1T$ in blue - see inset) for a junction in the (1, 1) (top) and (1 + 0.09, 1) (bottom) regimes as a function of the silicon gate voltage.
The simulations show that increasing the silicon gate voltage enhances the charge accumulation within the junction region in the $(1+0.09,1)$ regime, while the effect remains much weaker in the homogeneous $(1,1)$ configuration. This supports the interpretation that the silicon gate locally reinforces the electrostatic confinement responsible for the formation of the confined density pocket.
This corroborates our expectation of a 60-70nm distance between the side graphite gates.

\subsubsection{Physical interpretation of the confined skyrmion regime}
Overall, the self-consistent electrostatic simulations support the interpretation that the experimentally observed oscillatory regime originates from a confined skyrmionic phase localized at the junction.
In the experimentally relevant $(1+\epsilon,1)$ regime, the combination of the split graphite gates and the exposed silicon back gate produces a localized enhancement of the electronic density within the junction region.
The resulting charge-density pocket develops only for a restricted range of excess filling factors, with optimal confinement occurring near $\epsilon \approx 0.05-0.1$, while larger values progressively wash out the localized structure.
This establishes that the local filling factor inside the junction differs from that of the surrounding bulk quantum Hall ferromagnet, despite the left region remaining globally close to $\nu=1$.

The simulations further show that the spatial extent of the confined density enhancement is comparable to a few magnetic lengths $l_B$, which corresponds to the characteristic size expected for charged skyrmionic excitations.
This therefore suggests that the junction acts as an electrostatically-defined quasi-one-dimensional trapping potential capable of localizing skyrmions.
In contrast, skyrmions that may exist in the extended left region are expected to remain dilute and spatially distributed, leading primarily to smooth background scattering rather than discrete features in the non-local signal.

Importantly, the electrostatic reconstruction evolves continuously with gate voltage and therefore cannot by itself explain the robust quasi-periodic fluctuations observed experimentally.
Instead, the oscillatory regime appears only when the simulations predict the formation of the confined density pocket, supporting an interpretation in terms of discrete charging of localized skyrmionic excitations within the junction region.
Within this picture, the left graphite gate primarily controls the excess filling factor $\epsilon$ and therefore the number of available excess charges, while the silicon gate tunes the confinement potential itself by modifying the amplitude and spatial extent of the localized density enhancement.

Finally, we speculate that in the regime where $\epsilon$ maximizes the modulation of the electrostatic density bump, the resulting increase of the local filling-factor gradient may favor the stabilization of an additional row of localized skyrmions within the junction.
The intrinsic left--right asymmetry of the electrostatic confinement potential could then naturally explain why these additional localized states emerge at different gate voltages on the two sides of the junction.

We note that the simulations predict a confined density enhancement at the junction even for $\epsilon = 0$. However, the presence of an electrostatic confinement potential alone is not sufficient to nucleate skyrmions.
At $\nu = 1$, the quantum Hall ferromagnet remains incompressible and no excess charge is available to form charged skyrmionic excitations. In contrast, for $\epsilon > 0$, additional carriers can populate the confined density pocket, making localized skyrmionic textures energetically favorable.
The observed oscillatory regime therefore requires both electrostatic confinement and a finite excess filling away from the integer quantum Hall state.

\newpage
\clearpage
\section{Assessment of a possible quantum-dot origin of the time-dependent stochastic features}

It is important to consider whether the discrete features in Fig. 2D and 2E could arise from the formation of an accidental quantum dot near the constriction rather than from skyrmionic textures. In such a scenario, magnons—carrying an effective electric dipole moment—could couple to localized charges in a quantum dot, and gate-induced changes in the dot’s charge configuration could modify magnon scattering, leading to variations in the measured non-local voltage. While this mechanism is in principle possible, several quantitative and qualitative arguments make it unlikely.

First, the characteristic size required for such a dot would be comparable to the full constriction length, approximately 500 nm. For a dot of this scale, the expected gate-voltage period associated with Coulomb blockade can be estimated from the capacitance $C_g \approx \varepsilon_0 \varepsilon_r A / t$, where $A \sim \pi(250 \text{ nm})^2$ and a typical dielectric thickness is $t \sim 30–50$ nm. This yields $C_g \sim 10^{-16}$ F and a corresponding charging scale $\Delta V_g \sim e/C_g \sim 1$ mV. This scale is significantly smaller than the gate-voltage range over which the discrete features in Fig. 2D and 2E evolve. A conventional quantum dot of this size would therefore be expected to produce a much denser set of Coulomb-blockade oscillations than observed.

Second, accidental quantum dots typically give rise to irregular, strongly sample-dependent Coulomb-blockade fluctuations that are highly sensitive to disorder, local charge rearrangements, and thermal cycling. In contrast, the observed features are reproducible, exhibit quasi-periodicity, and appear specifically in the narrow filling-factor range near $\nu \approx 1$, where skyrmion formation is expected. Their confinement to the constriction is more naturally explained by the nucleation and rearrangement of skyrmionic textures in a confined electrostatic landscape than by an uncontrolled disorder-induced dot.

Finally, the observed time-dependent noise correlations are not naturally captured by a static quantum-dot picture. Whereas a quantum dot would primarily affect transport through discrete changes in charge occupation, the measured fluctuations are more consistent with slow collective rearrangements of an interacting many-body state, as expected for a weakly pinned or dynamically fluctuating skyrmion configuration.

We therefore cannot fully exclude the presence of local charge inhomogeneities near the constriction. However, a conventional accidental quantum dot of order 500 nm is inconsistent with both the observed gate-voltage scale and the reproducibility and selectivity of the features. Overall, a quantum-dot origin is strongly disfavored compared to the skyrmion-based interpretation.

\clearpage

%% file: Extended_data.tex
\title{Extended Data for: Magnons reveal topology and dynamics of a skyrmion crystal}

\author{Rapha\"el Ayache}
\thanks{These authors contributed equally}
\affiliation{SPEC, CEA, CNRS, Université Paris-Saclay, CEA Saclay, 91191 Gif sur Yvette Cedex
France}

\author{Nilotpal Chakraborti}
\thanks{These authors contributed equally}
\affiliation{TCM Group, Cavendish Laboratory, University of Cambridge, Cambridge CB3 0HE, United Kingdom}

\author{Manabendra Kuiri}
\thanks{These authors contributed equally}
\affiliation{SPEC, CEA, CNRS, Université Paris-Saclay, CEA Saclay, 91191 Gif sur Yvette Cedex
France}
\affiliation{Department of Physics, Birla Institute of Technology and Science, Pilani, Hyderabad Campus, Jawahar Nagar, Kapra Mandal, Medchal District, Telangana 500078, India}

\author{Quentin Benichou}
\affiliation{SPEC, CEA, CNRS, Université Paris-Saclay, CEA Saclay, 91191 Gif sur Yvette Cedex
France}

\author{Antonio Lacerda-Santos}
\affiliation{SPEC, CEA, CNRS, Université Paris-Saclay, CEA Saclay, 91191 Gif sur Yvette Cedex
France}

\author{Lilian Seyve}
\affiliation{SPEC, CEA, CNRS, Université Paris-Saclay, CEA Saclay, 91191 Gif sur Yvette Cedex
France}

\author{Himadri Chakraborti}
\affiliation{SPEC, CEA, CNRS, Université Paris-Saclay, CEA Saclay, 91191 Gif sur Yvette Cedex
France}
\affiliation{Laboratory of Atomic and Solid State Physics, Cornell University, Ithaca 14850, NY, USA}

\author{L\'eo Pugliese}
\affiliation{SPEC, CEA, CNRS, Université Paris-Saclay, CEA Saclay, 91191 Gif sur Yvette Cedex
France}

\author{Kenji Watanabe}
\affiliation{Research Center for Functional Materials, National Institute for Materials Science, Japan}

\author{Takashi Taniguchi}
\affiliation{International Center for Materials Nanoarchitectonics, National Institute for Materials Science, Japan}

\author{Roderich Moessner}
\affiliation{Max-Planck Institut für Physik komplexer Systeme, Nöthnitzer Straße 38, Dresden 01187, Germany}

\author{Cosimo Gorini}
\affiliation{SPEC, CEA, CNRS, Université Paris-Saclay, CEA Saclay, 91191 Gif sur Yvette Cedex
France}

\author{Beno\^it Doucot}
\affiliation{LPTHE, UMR 7589, CNRS and Sorbonne Université, 75252 Paris Cedex 05, France}

\author{Preden Roulleau}
\email{preden.roulleau@cea.fr}
\affiliation{SPEC, CEA, CNRS, Université Paris-Saclay, CEA Saclay, 91191 Gif sur Yvette Cedex
France}

\maketitle

\renewcommand{\thefigure}{E.D.\arabic{figure}}
\begin{figure*}[h]
    \centering
    \includegraphics[width=.8\textwidth]{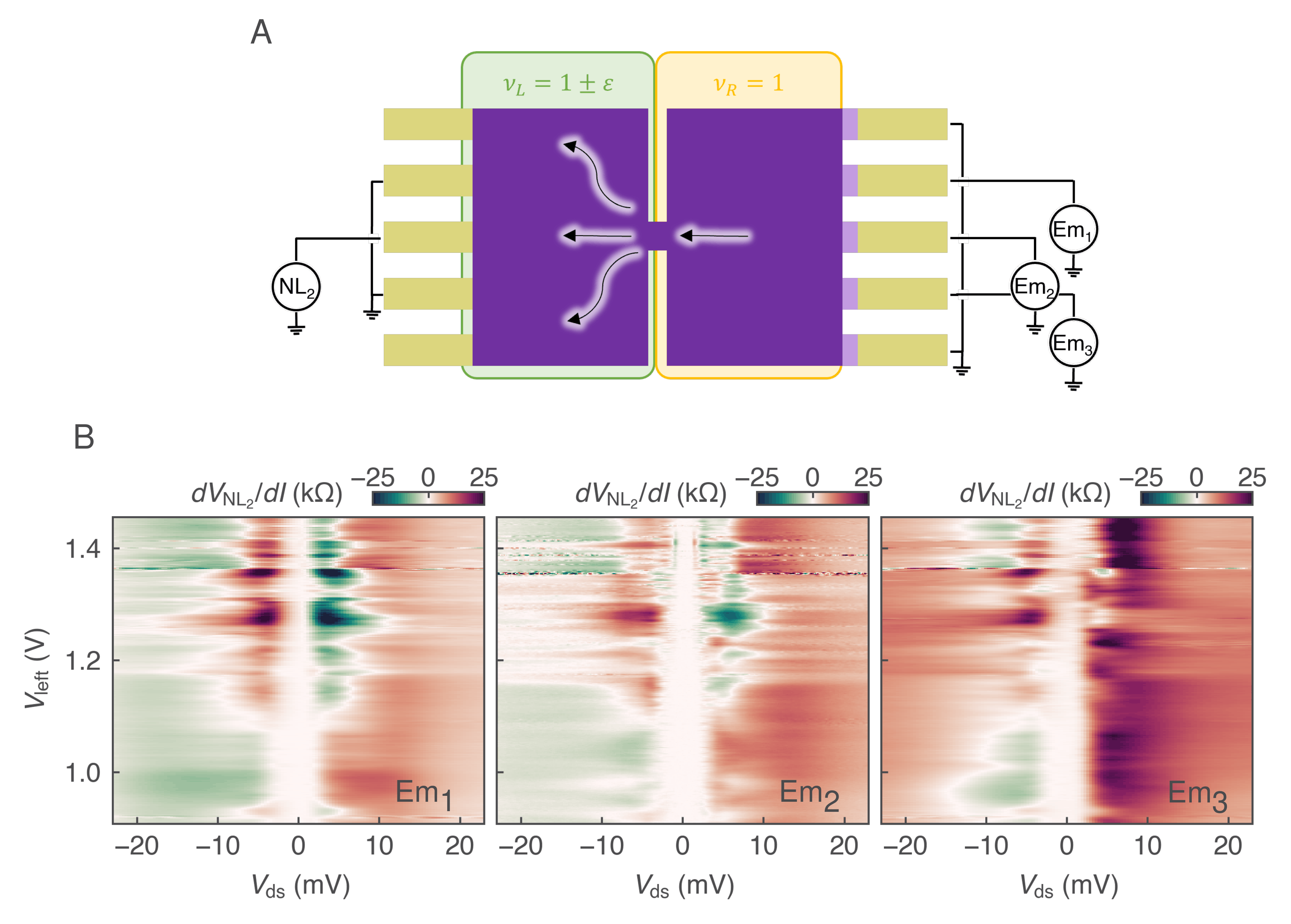}
    \caption{
        \textbf{(a)} Schematic of the experimental geometry showing the three magnon emitters, $E_{\mathrm{m,1}}$, $E_{\mathrm{m,2}}$, and $E_{\mathrm{m,3}}$, positioned at different lateral angles with respect to the gated junction region. Magnons emitted from these contacts propagate toward the junction, where they interact with the in-plane electric field and experience momentum-dependent scattering before reaching the nonlocal detector NL2.
        \textbf{(b)} Nonlocal magnon signals $dV_{NL}/dI$ measured at NL2 for each of the three emitters. The lateral emitters produce a substantially stronger response due to their larger transverse momentum components, which enhance the coupling between the magnon electric dipole and the in-plane electric field, leading to increased scattering and redirection toward NL2. In contrast, magnons injected from the central emitter $E_{\mathrm{m,2}}$ predominantly follow near-normal trajectories across the junction and couple more weakly to the electric field, resulting in a comparatively smaller nonlocal signal.
    }
    \label{figure_moment}
\end{figure*}

\begin{figure*}[t]
    \centering
    \includegraphics[width=.8\textwidth]{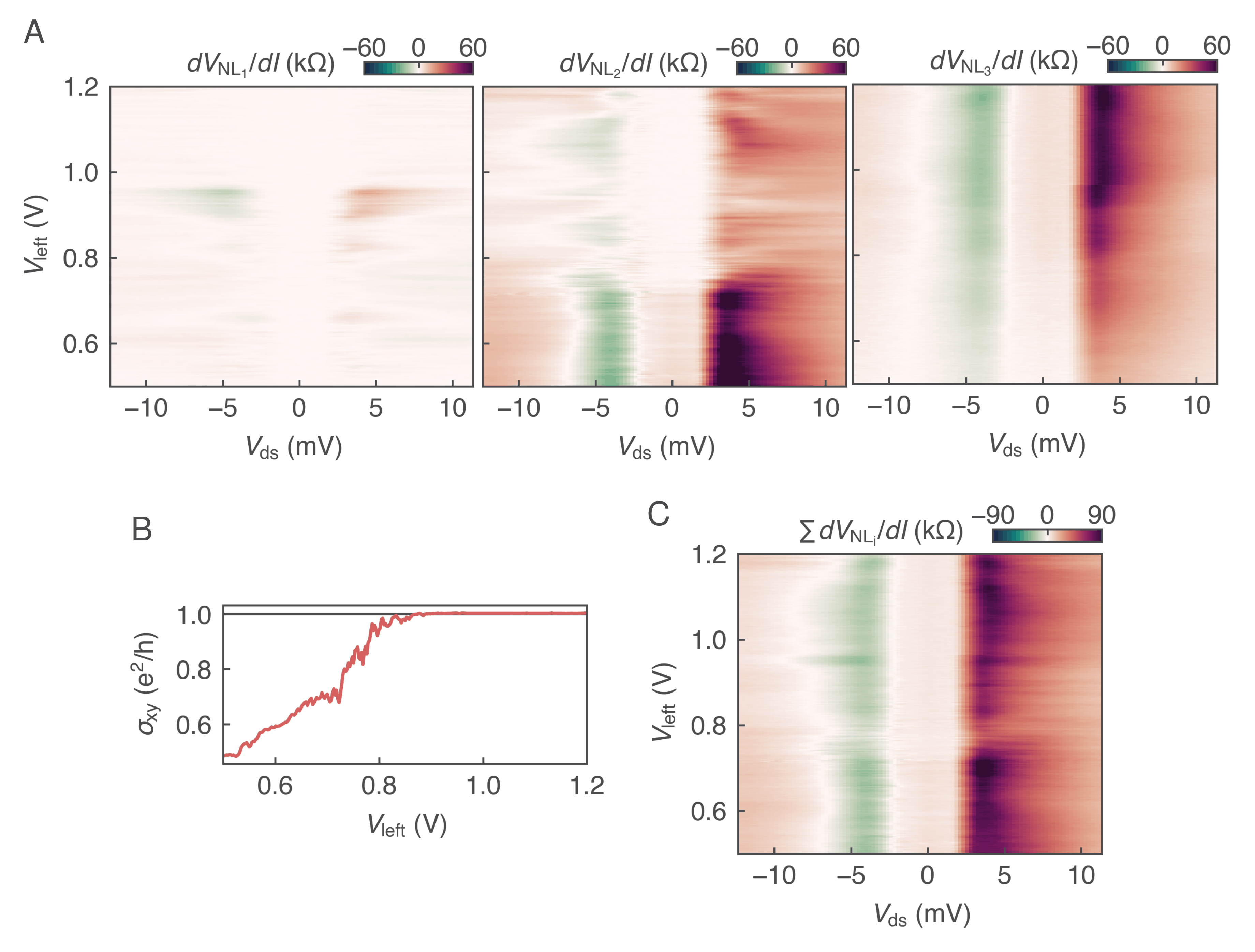}
    \caption{\textbf{regime $\mathbf{(1-\varepsilon, 1)}$}:
        \textbf{(a)} Non-local magnon signals measured at detectors NL1, NL2, and NL3 while exciting the central emitter $E_{\mathrm{m,2}}$. Each trace corresponds to a single detector, showing the distribution of the magnon signal across the device.
        \textbf{(b)} Simultaneously measured Hall conductivity $\sigma_{xy}$ in the right-gated region.
        \textbf{(c)} Total integrated magnon signal summed over all three detectors. The signal remains approximately constant across the range of carrier densities studied, demonstrating conservation of the magnon flux through the bulk despite small variations in individual detector amplitudes. No distinctive skyrmion-crystal signatures are observed from this measurement perspective.
    }
    \label{fig:magnon_conservation}
\end{figure*}

\begin{figure}
    \centering
    \includegraphics[width = 1\textwidth]{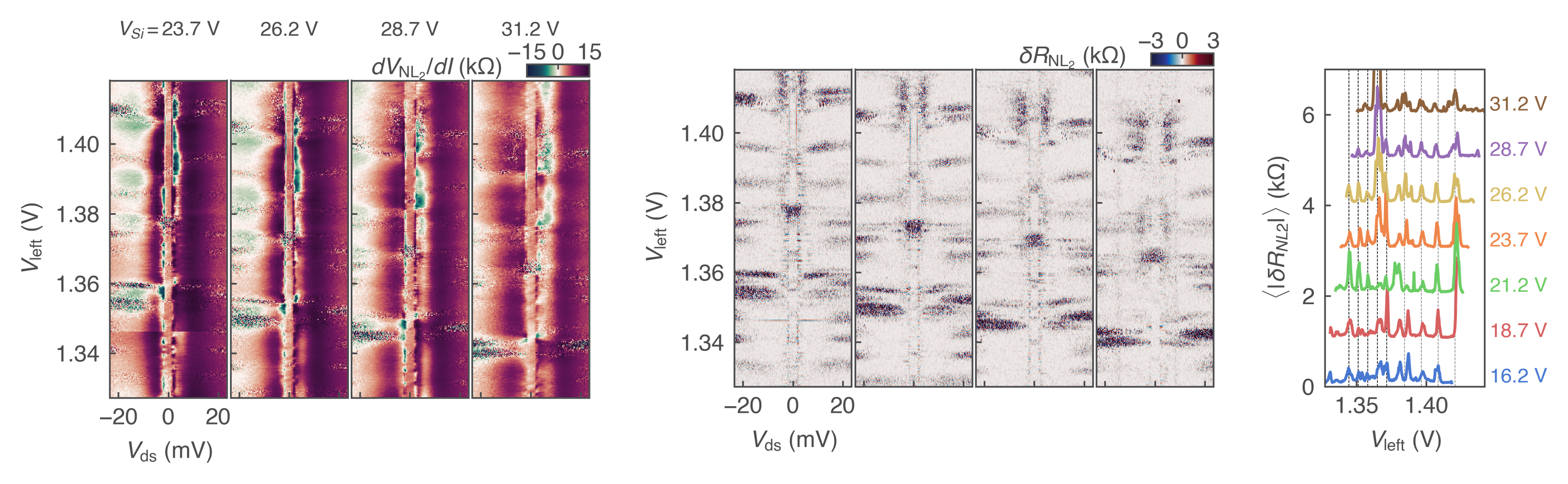}
    \caption{(a–b) Data from Fig.~2D and 2E without guide lines. (c) Bias-averaged (for $|V_{\mathrm{dc}}| > 10$ mV) absolute value of the background-subtracted non-local resistance $\delta R_{\mathrm{NL2}}$ as a function of $V_{\mathrm{left}}$ for different silicon gate voltages (indicated on the right of the plot). The black dashed lines are spaced by 6.8 mV and the grey lines by 12 mV. For clarity, the curves have been horizontally shifted to align the oscillations and highlight their common periodicity across the different silicon gate voltages. These results demonstrate a quasi-periodic dependence on left gate voltage of the noisy features, which can be tuned via the silicon gate acting only on the junction region.}
\end{figure}

\begin{figure}[h]
    \centering
    \includegraphics[width=0.5\textwidth]{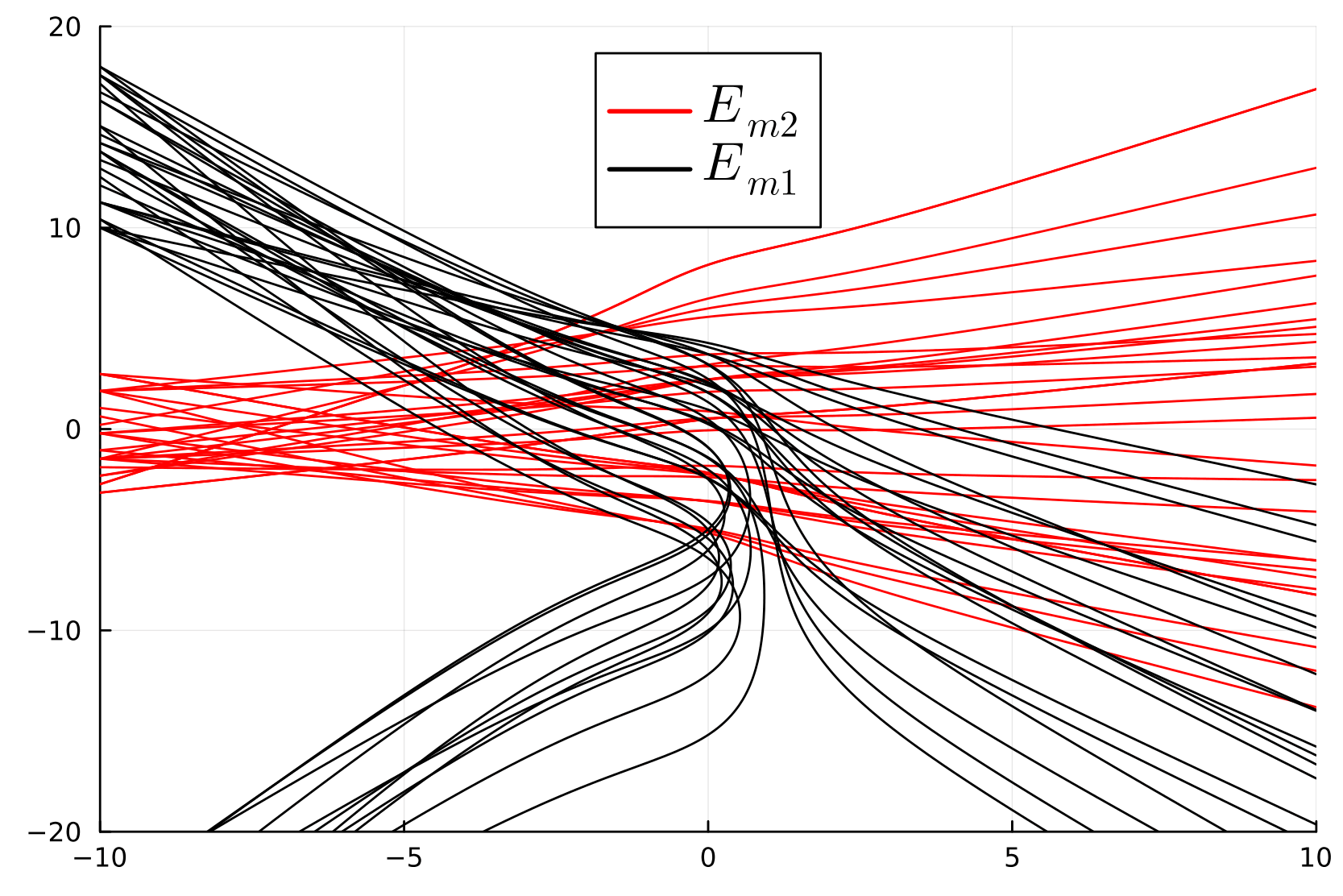}
    \caption{Classical magnon trajectories in the presence of an electric field
    parallel to the horizontal axis, independent of $y$, and confined to a central region of finite width. As expected from our general discussion, red trajectories with a small value  of the transverse momentum $p_y$ are only weakly deflected, with a consistently small, negative vertical shift $\Delta y$. By contrast,
    when the absolute value of $p_y$ increases (black trajectories), the vertical shift increases in magnitude, and tends to diverge for the critical value $p_{y*}(p_x)$ separating transmitted trajectories and bouncing ones.
    }
    \label{fig:magnon_deflection}
\end{figure}

\begin{figure}
    \centering
    \includegraphics[width = 0.8\textwidth]{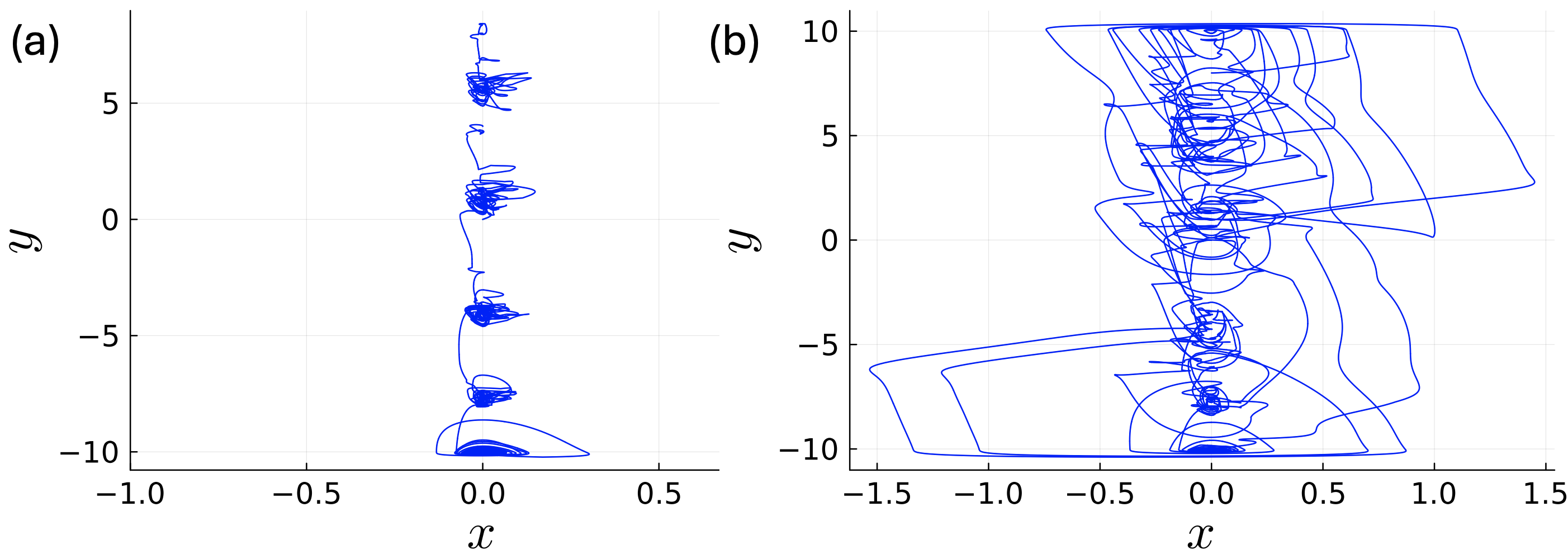}
    \caption{Skyrmion centre positions for high (a) and low (b) stiffness altered through shaking by magnons. We see that for different values of $K_x$, the degree to which the impinging magnons alter the collective motion of the crystal varies drastically.  }
    \label{skyrmcoords}
\end{figure}